\PassOptionsToPackage{dvipsnames,table}{xcolor}
\documentclass[sigconf]{acmart}
\usepackage{amsmath}
\usepackage{subcaption}
\usepackage{algorithm}
\usepackage{algpseudocode}
\usepackage{xspace}
\usepackage{filecontents}
\usepackage{wrapfig}
\usepackage{tikz}
\usepackage{booktabs,tabularx}
\newcommand*\circled[1]{\tikz[baseline=(char.base)]{
            \node[shape=circle,draw,inner sep=2pt] (char) {#1};}}

\AtBeginDocument{%
  }

\acmYear{2025}\copyrightyear{2025}
\setcopyright{acmlicensed}
\acmConference[SoCC '25]{ACM Symposium on Cloud Computing}{November 19--21, 2025}{Online, USA}
\acmBooktitle{ACM Symposium on Cloud Computing (SoCC '25), November 19--21, 2025, Online, USA}
\acmDOI{10.1145/3772052.3772208}
\acmISBN{979-8-4007-2276-9/25/11}

\begin{document}

\def\name{DFUSE\xspace}
\title{\name: Strongly Consistent Write-Back Kernel Caching for Distributed Userspace File Systems}

\author{
{\rm Haoyu Li}$^\dag$ \quad
{\rm Jingkai Fu}$^\dag$ \quad
{\rm Qing Li}$^\natural$ \quad
{\rm Windsor Hsu}$^\natural$ \quad
{\rm Asaf Cidon}$^\dag$ \\
Columbia University$^\dag$ \quad\quad Alibaba Cloud$^\natural$
}

\renewcommand{\shortauthors}{Li et al.}

\begin{abstract}
Cloud platforms host thousands of tenants that demand POSIX semantics, high throughput, and rapid evolution from their storage layer. Kernel-native distributed file systems supply raw speed, but their privileged code base couples every release to the kernel, widens the blast radius of crashes, and slows innovation. FUSE-based distributed file systems flip those trade-offs: they run in user space for fast deployment and strong fault isolation, yet the FUSE interface disables the kernel's write-back page cache whenever strong consistency is required. Practitioners must therefore choose between (i) weak consistency with fast write-back caching or (ii) strong consistency with slow write-through I/O, a limitation that has kept FUSE distributed file systems out of write-intensive cloud workloads.

To this end, we present \name, the first distributed FUSE file system that delivers write-back kernel caching and strong consistency. \name achieves this by offloading userspace consistency control to the kernel driver, allowing coordinated access to the kernel's page cache across nodes. This design eliminates blind local cache updates and ensures cluster-wide strong consistency without compromising performance. In our evaluation, \name achieves up to 68.0\% higher throughput and 40.4\% lower latency than the existing write-through design of FUSE-based distributed file systems.
\end{abstract}




\maketitle

\section{Introduction}

\begin{table*}[t]
  \caption{Comparison between \name\ and existing solutions.}
  \label{tab:comparison}
  \centering
  \begin{tabularx}{\textwidth}{@{}l X X X X X@{}}
    \toprule
    &                           
    &                           
    \multicolumn{2}{c}{\textbf{Existing FUSE-based DFSes}} &
    \multicolumn{2}{c}{\textbf{Existing Kernel-native DFSes}} \\
    \cmidrule(lr){3-4}\cmidrule(lr){5-6}
    \textbf{Dimension} & \textbf{\name} & \textbf{Ceph-fuse} & \textbf{GlusterFS}
                       & \textbf{Lustre} & \textbf{GPFS} \\
    \midrule
    Client location        & \textcolor{ForestGreen}{User space}
                           & \textcolor{ForestGreen}{User space}
                           & \textcolor{ForestGreen}{User space}
                           & \textcolor{BrickRed}{Kernel}
                           & \textcolor{BrickRed}{Kernel} \\
    \hline
    Page-cache mode        & \textcolor{ForestGreen}{Write-back}
                           & \textcolor{BrickRed}{Write-through (slow)}
                           & \textcolor{ForestGreen}{Write-back}
                           & \textcolor{ForestGreen}{Write-back}
                           & \textcolor{ForestGreen}{Write-back} \\
    \hline
    Consistency            & \textcolor{ForestGreen}{Strong} 
                           & \textcolor{ForestGreen}{Strong}
                           & \textcolor{BrickRed}{Weak}
                           & \textcolor{ForestGreen}{Strong}
                           & \textcolor{ForestGreen}{Strong}\\
    \hline
    Fault isolation        & \textcolor{ForestGreen}{Strong}
                           & \textcolor{ForestGreen}{Strong}
                           & \textcolor{ForestGreen}{Strong}
                           & \textcolor{BrickRed}{Weak}
                           & \textcolor{BrickRed}{Weak} \\
    \hline
    Deployment             & \textcolor{ForestGreen}{Unprivileged}
                           & \textcolor{orange}{Mostly unprivileged}
                           & \textcolor{ForestGreen}{Unprivileged}
                           & \textcolor{BrickRed}{Privileged}
                           & \textcolor{BrickRed}{Privileged} \\
    \hline
    Development            & \textcolor{ForestGreen}{Fast}  
                           & \textcolor{ForestGreen}{Fast}
                           & \textcolor{ForestGreen}{Fast}
                           & \textcolor{BrickRed}{Slow}
                           & \textcolor{BrickRed}{Slow} \\
    \bottomrule
  \end{tabularx}
\end{table*}

Large, multi-tenant clouds demand storage that can evolve quickly, survive component failures, and plug into unmodified POSIX applications~\cite{weil2006ceph, rajgarhia2010fuse, noauthor_cloud_nodate-1, noauthor_cloud_nodate}. Kernel-native distributed file systems~\cite{schwan2003lustre, schmuck2002gpfs} excel at raw throughput, but their code runs with full kernel privileges: a single bug can crash the entire node, upgrades require kernel rollouts, and every new feature must wait for the next kernel release cycle. These constraints slow innovation and increase the blast radius of faults (Table~\ref{tab:comparison}). To sidestep those hurdles, cloud operators are increasingly embracing FUSE-based DFSs—userspace file systems that preserve the kernel's VFS and page-cache interfaces while isolating file-system logic in separate processes. FUSE lets developers iterate rapidly, ship features without kernel changes, and contain crashes to the file-system process itself, making it a natural fit for fast-moving services such as object-store gateways, snapshot engines, and ML-training checkpoints \cite{noauthor_cloud_nodate, noauthor_deepseek-ai3fs_2025, noauthor_s3fs-fuses3fs-fuse_2025, noauthor_juicefs_nodate, noauthor_miniominfs_2025, cheung_kahinggoofys_2025, noauthor_use_nodate, noauthor_shared_nodate}. Recent innovations \cite{huai2021xfuse, cho2024rfuse, zhu2018direct, bijlani2019extension, ren2013tablefs} further mitigate historical performance bottlenecks making FUSE increasingly viable for large-scale deployments.  

However, a fundamental challenge persists: FUSE-based distributed file systems struggle to provide both high performance and strong consistency (i.e., allow distributed clients to read the most recent written value). The kernel's page cache—a key driver of efficiency in local file systems—remains largely incompatible with distributed environments. The write-back page cache enables applications to return from I/O operations immediately after buffering data in the kernel, removing the expensive userspace synchronization of FUSE from the critical path. Yet, in distributed settings, this risks data races and inconsistency, as nodes may operate on stale or uncoordinated cached copies. For example, if node A writes data to its local page cache (happens in the kernel) without immediately synchronizing with other nodes (happens in the userspace), node B might read stale data or make conflicting updates, leading to data corruption. Using the write-through mode of kernel page cache mitigates the consistency issue at the cost of low performance (Figure~\ref{fig:fuse-breadkdown}). Therefore, prior solutions either sacrifice strong consistency for performance (e.g., enabling a write-back cache without synchronization)\cite{noauthor_performance_2011, noauthor_gluster_nodate} or default to write-through caching\cite{weil2006ceph}, which synchronizes every write to userspace (Table~\ref{tab:comparison}). In write-through mode, each operation must complete synchronization before returning to the application, incurring high latency. This trade-off limits practicality in consistency-sensitive cloud workloads, such as databases or real-time analytics.  

We present \name, a FUSE-based distributed file system that bridges the gap between high performance and strong consistency. \name enables write-back page caching while enforcing strong consistency across nodes, eliminating the need to compromise between performance and correctness. The core insight of \name is offloading userspace consistency control (e.g., lease management) to the kernel FUSE driver. By embedding distributed metadata directly into the FUSE driver, \name coordinates access permissions within the kernel. This design resolves the kernel space deadlock issue in current systems (discussed in \S\ref{sec:challenge-lock}). This allows nodes to gain cluster-wide access within the kernel during I/O operations, ensuring that only authorized nodes read or modify cached data. For instance, when a write lease is revoked, the kernel proactively invalidates the local page cache and flushes dirty pages. This design retains FUSE's userspace flexibility while leveraging the kernel's efficiency, avoiding costly round trips for common-case operations.  

Our evaluation demonstrates that \name consistently improves I/O throughput and latencies compared to a baseline, which represents the architecture of systems such as GlusterFS\cite{noauthor_gluster_nodate}, an open-source distributed file system based on FUSE. Results demonstrate a 1.75x throughput improvement at best, outperforming write-through approaches. \name also scales linearly with node count, maintaining consistency even at 25\% contention. 

Our contributions include:

\begin{itemize}
\item An analysis showing that leveraging kernel's page cache fundamentally breaks strong consistency in FUSE-based distributed file systems.
\item The kernel offloading mechanism that embeds distributed coordination logic into the FUSE driver.
\item The design and implementation of \name, the first FUSE-based system to support write-back caching with strong consistency.
\end{itemize}

The rest of this paper is organized as follows: \S\ref{sec:background} provides background on FUSE and FUSE-based distributed file systems; \S\ref{sec:challenges} details consistency and locking challenges; \S\ref{sec:design} and \S\ref{sec:impl} present \name's design and implementation, respectively; \S\ref{sec:eval} evaluates its efficacy; \S\ref{sec:related} discusses related work; and \S\ref{sec:conclusion} concludes the paper.
\section{Background}\label{sec:background}
\subsection{FUSE in Cloud Computing}\label{sec:fuse-in-cloud}
FUSE (Filesystem in Userspace) has become a very popular file system for modern cloud infrastructure~\cite{noauthor_cloud_nodate, noauthor_deepseek-ai3fs_2025, noauthor_s3fs-fuses3fs-fuse_2025, noauthor_juicefs_nodate, noauthor_miniominfs_2025, cheung_kahinggoofys_2025, noauthor_use_nodate, noauthor_shared_nodate}, enabling developers to build custom file systems without modifying the kernel. By running the file system in userspace, FUSE prevents kernel crashes due to bugs and accelerates development cycles. For example, systems like S3FS~\cite{noauthor_s3fs-fuses3fs-fuse_2025} (mounting cloud object storage as POSIX file systems) and JuiceFS~\cite{noauthor_juicefs_nodate} (optimizing cloud data access with caching) leverage FUSE to expose a familiar POSIX and file system to applications while abstracting different backend storage interfaces such as object stores.

Recent projects further enhance FUSE's viability for large-scale deployments. For example, XFUSE~\cite{huai2021xfuse} reduces kernel-userspace context switches by batching requests and scalable communication patterns. Stackable file systems~\cite{ren2013tablefs} layer caching or encryption on top of existing FUSE implementations, improving metadata operation efficiency. These enhancements improve FUSE performance when running a file system on a single node, but fundamental challenges remain when trying to run a FUSE-based distributed file system.

\subsection{FUSE Architecture Overview}\label{sec:fuse}
\begin{figure}
    \centering
    \includegraphics[width=0.9\linewidth]{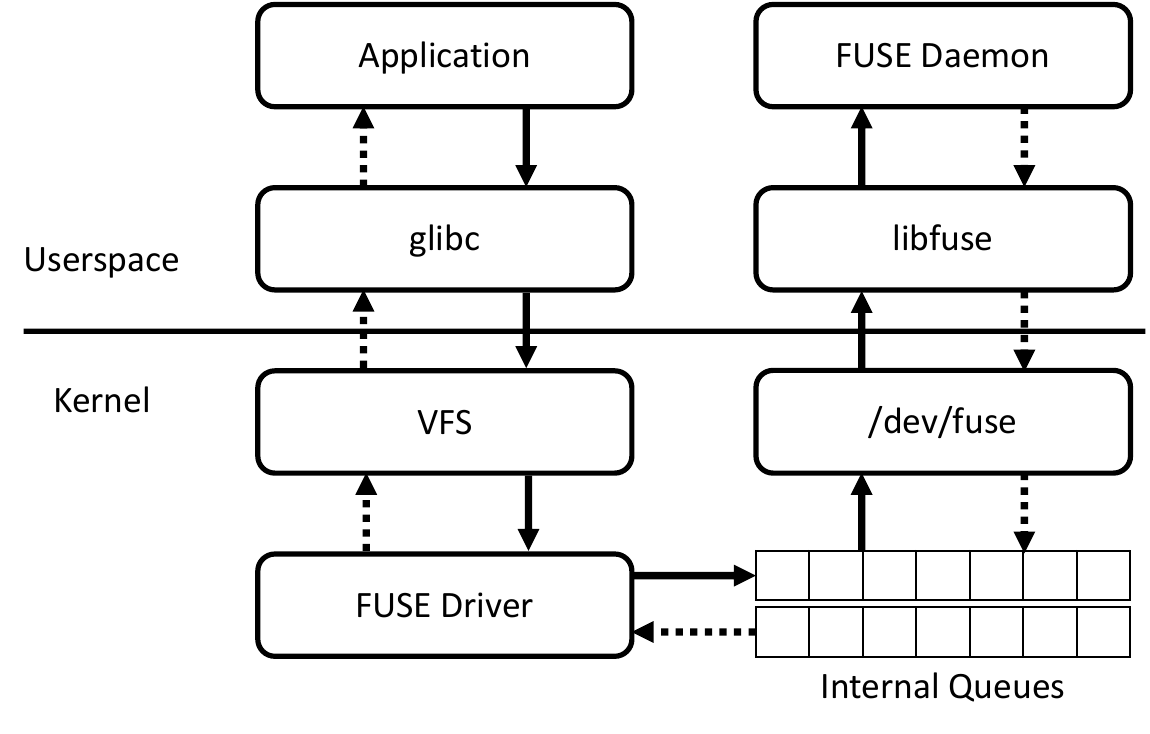}
    \caption{Overview of the FUSE Framework}
    \label{fig:fuse}
\end{figure}

FUSE is a framework that allows developers to build userspace file systems exportable to the Linux kernel. As illustrated in Fig~\ref{fig:fuse}, it comprises three key components: the FUSE kernel driver, the userspace library (libfuse~\cite{noauthor_libfuselibfuse_2025}), and a custom FUSE daemon. The kernel driver registers the file system with the VFS, allowing applications to interact with it via standard POSIX interfaces (e.g., open, read, write) exposed by the GNU C Library (glibc). When an application issues a syscall, glibc invokes the corresponding VFS function, which routes the request to the FUSE driver. The driver translates the syscall into a FUSE protocol message (e.g., \texttt{FUSE\_READ}) and forwards it to userspace via a special device called /dev/fuse. To summarize, while FUSE is a userspace-based file system, when applications interact with it, the I/O request gets routed first to the kernel then back to the file system in userspace.

The userspace library libfuse abstracts low-level protocol details, providing developers with higher-level APIs to implement file system operations (e.g., read, write) in the custom FUSE daemon. This daemon—a standalone process—executes developer-defined logic, such as fetching data from remote storage or enforcing access policies. Once the daemon processes a request, it returns a response through /dev/fuse to the kernel driver, which propagates the result back to the application via the VFS. 

Communication between the kernel driver and userspace daemon occurs through /dev/fuse, which maintains prioritized queues for different request types. The daemon polls these queues, blocking when empty until the driver signals new requests. Moreover, FUSE supports a notification mechanism: the daemon can proactively notify the kernel driver of certain events. For example, in distributed scenarios, upon receiving a revocation request from a lock manager, the daemon can notify the kernel driver to flush buffered data from the page cache.

The above FUSE's architectural design introduces performance overheads, primarily due to frequent kernel-userspace mode switches and extra memory copies~\cite{vangoor2017fuse, vangoor2019performance, zhu2018direct, huai2021xfuse, cho2024rfuse, kappes2024faster}, where every file system operation traverses the kernel-to-userspace boundary via the /dev/fuse device. For instance, a simple \texttt{write()} syscall incurs multiple context switches (first to the kernel driver, then to the userspace daemon, and back) along with data copies between kernel and userspace buffers.

\subsection{Leveraging the Page Cache}
Even though the file system implementation resides in userspace, the FUSE kernel driver utilizes the kernel's page cache to accelerate I/O operations by indexing cached data using inode numbers and page offsets.
FUSE leverages the kernel's page cache in one of two modes.  In write-back mode, writes return immediately after buffering data in the kernel cache, deferring synchronization to the userspace daemon. This minimizes both mode switches (by avoiding synchronous userspace round trips) and memory copies (by retaining data in the kernel's page cache). Consequently, write-back mode significantly reduces latency and improves throughput, making it attractive for performance-sensitive workloads.

In contrast, write-through mode synchronously updates both the kernel cache and userspace daemon. Then, the userspace daemon can propagate the update to other nodes, ensuring strong consistency in a distributed file system setting. However, this approach increases the write latency as each write has to go to userspace (the daemon typically maintains its own userspace in-memory cache to reduce communication overhead with remote storage services in distributed FUSE-based file systems).

\subsection{FUSE Performance Breakdown}
\begin{figure}
    \centering
    \includegraphics[width=1\linewidth]{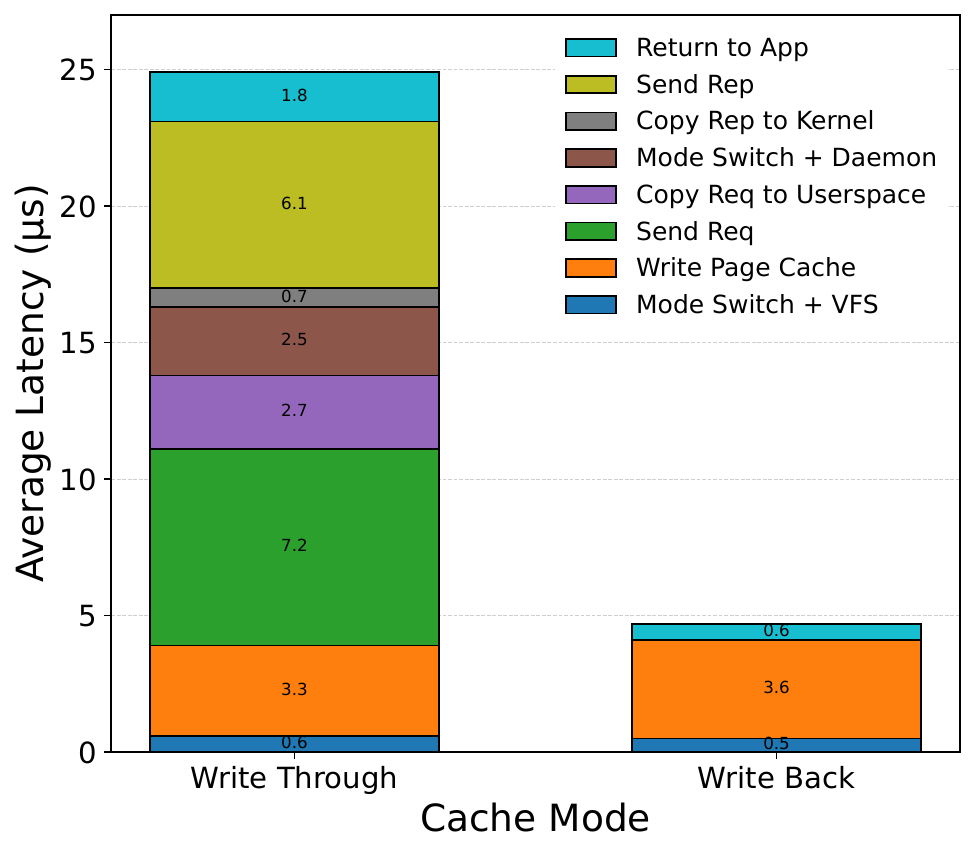}
    \caption{Latency breakdown of write requests in FUSE}
    \label{fig:fuse-breadkdown}
\end{figure}

We quantify the performance benefits of kernel's write-back page cache in Fig~\ref{fig:fuse-breadkdown}. To break down the overhead of the FUSE protocol itself, we implement a \texttt{NullFS} which simply returns zero in any file system operations in the userspace. Note that none of the requests actually go to the disk. The stacked bars (from the bottom up) present the life cycles of write requests chronologically.\footnote{The workload is 1M random 4KB writes over 1K files of 16MB after warm-up.} First, the application issues a write system call, which triggers a mode switch into the kernel, and the VFS layer dispatches the request to the FUSE driver. In both cache modes, the content within the write request will be copied to the kernel's page cache. If the write-back cache is enabled, FUSE then returns the number of written bytes or an error to the application, with an average total latency of 4.7 $\mu$s (the sum of the left bar). 

In contrast, in the write-through mode, after writing the page cache, FUSE continues to pass the write request to the userspace daemon. In particular, the FUSE driver as in Fig~\ref{fig:fuse} first places the request in one of the internal queues and next wakes up the daemon thread that waits on the queues (7.2 $\mu$s). Subsequently, the daemon thread dequeues the request and copies it to the userspace buffer (2.7 $\mu$s). After switching to the userspace, the daemon invokes the userspace write function to handle the requests (2.5 $\mu$s). Finally, the response returns to the kernel driver in the same way, i.e., the daemon copies the reply to the kernel buffer (0.7 $\mu$s) and notifies the FUSE driver thread of the incoming result (6.1 $\mu$s). 

From Fig~\ref{fig:fuse-breadkdown}, the overhead of the extra request round trip in FUSE write-through cache is on average 19.2 $\mu$s, while write-back cache mode incurs only 4.7 $\mu$s. 

\subsection{Achieving Strong Consistency}
Strong consistency is a critical property for distributed file systems, ensuring that any update made to data is immediately visible to subsequent read operations across all nodes. Implementing strong consistency typically involves synchronization mechanisms that ensure updates propagate immediately and atomically across all nodes. 

A common approach used in distributed file systems is the lease mechanism~\cite{gray1989leases, schmuck2002gpfs,schwan2003lustre,kronenberg1986vaxcluster,mohindra1994distributed}. Compared to other synchronization mechanisms, e.g., distributed locks, leases achieve lower synchronization overhead as the lease-holding node can safely perform operations locally during a lease period without communication with the lease coordinator. Once the lease expires or is revoked, the node must either renew it or cease operations on the leased data until it obtains a new lease.

In FUSE-based distributed file systems, implementing leases is particularly challenging due to the separation between kernel-resident page caches and userspace consistency logic. For instance, a node holding a write lease may buffer updates in its kernel page cache, leaving other nodes unaware of pending changes until a flush occurs. Prior systems like Ceph-FUSE resort to write-through caching to synchronize every update with userspace, while Gluster sacrifices consistency for performance. We elaborate on these challenges in \S\ref{sec:challenges} and explain how \name addresses these limitations by embedding lease management directly into the kernel driver, enabling write-back caching without compromising consistency in \S\ref{sec:design}.
\section{Challenges}\label{sec:challenges}
This section explores why it is challenging to leverage the kernel's page cache under either write-back or write-through mode effectively in distributed FUSE settings.

\begin{figure}[t!]
    \centering
    \begin{subfigure}[t]{0.45\textwidth}
        \centering
        \includegraphics[width=\textwidth]{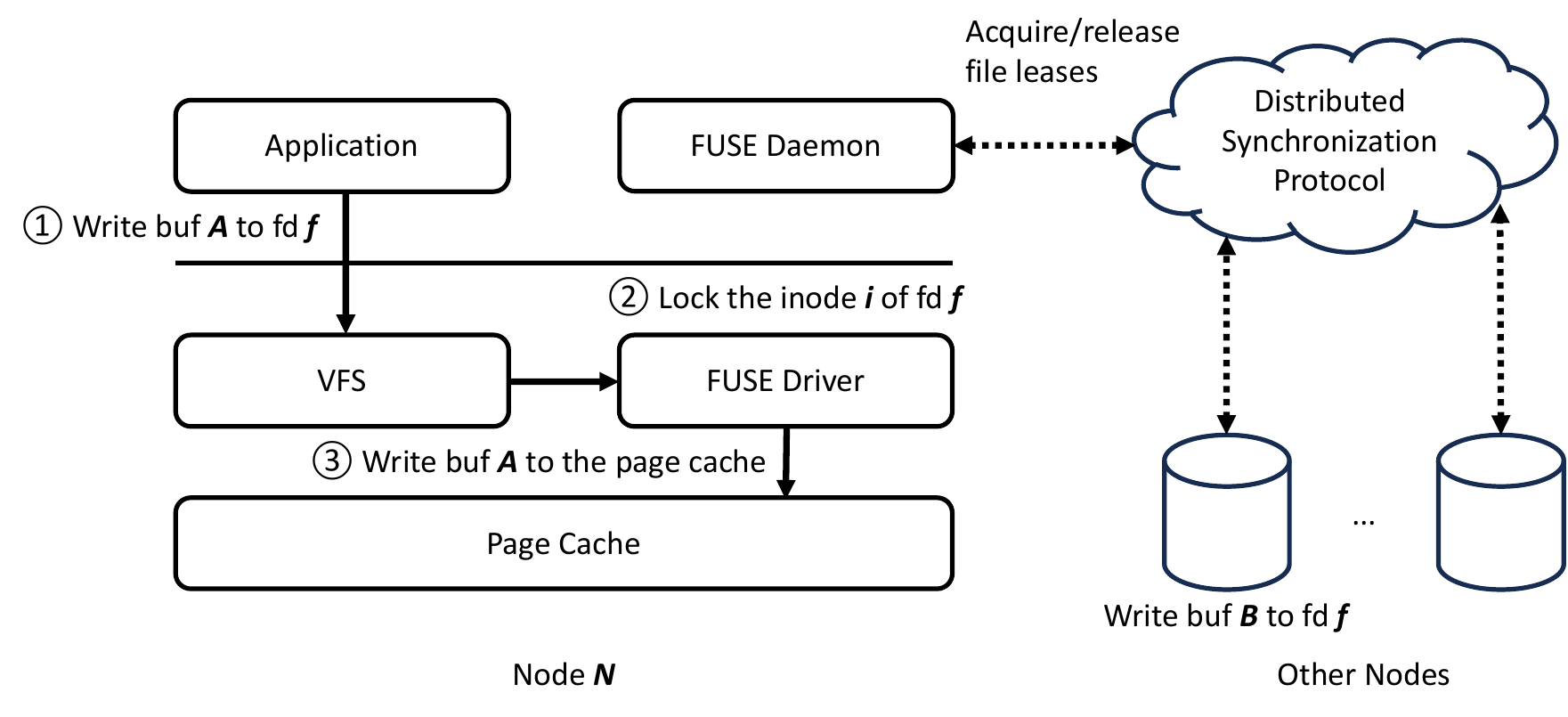}
        \caption{Data inconsistency caused by missing cross-layer coordination: kernel inode locks vs. userspace distributed protocols}
        \label{fig:write-back-eg}
    \end{subfigure}
    \hfill
    \begin{subfigure}[t]{0.45\textwidth}
        \centering
        \includegraphics[width=\textwidth]{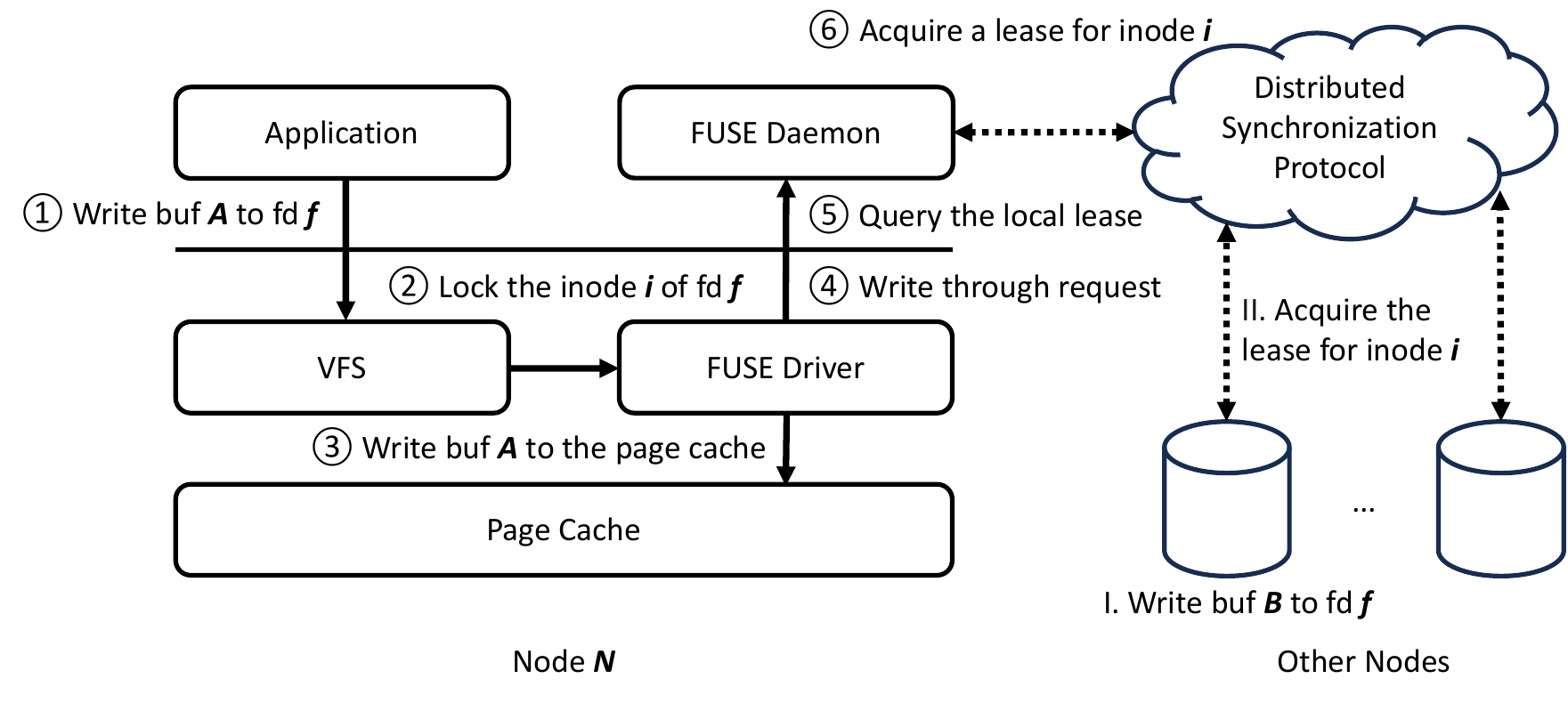}
        \caption{The workflow in a FUSE-based distributed file system. Each read and write request need to go through the userspace to obey the distributed synchronization protocol.}
        \label{fig:write-through-eg}
    \end{subfigure}
    \hfill
    \begin{subfigure}[t]{0.45\textwidth}
        \centering
        \includegraphics[width=\textwidth]{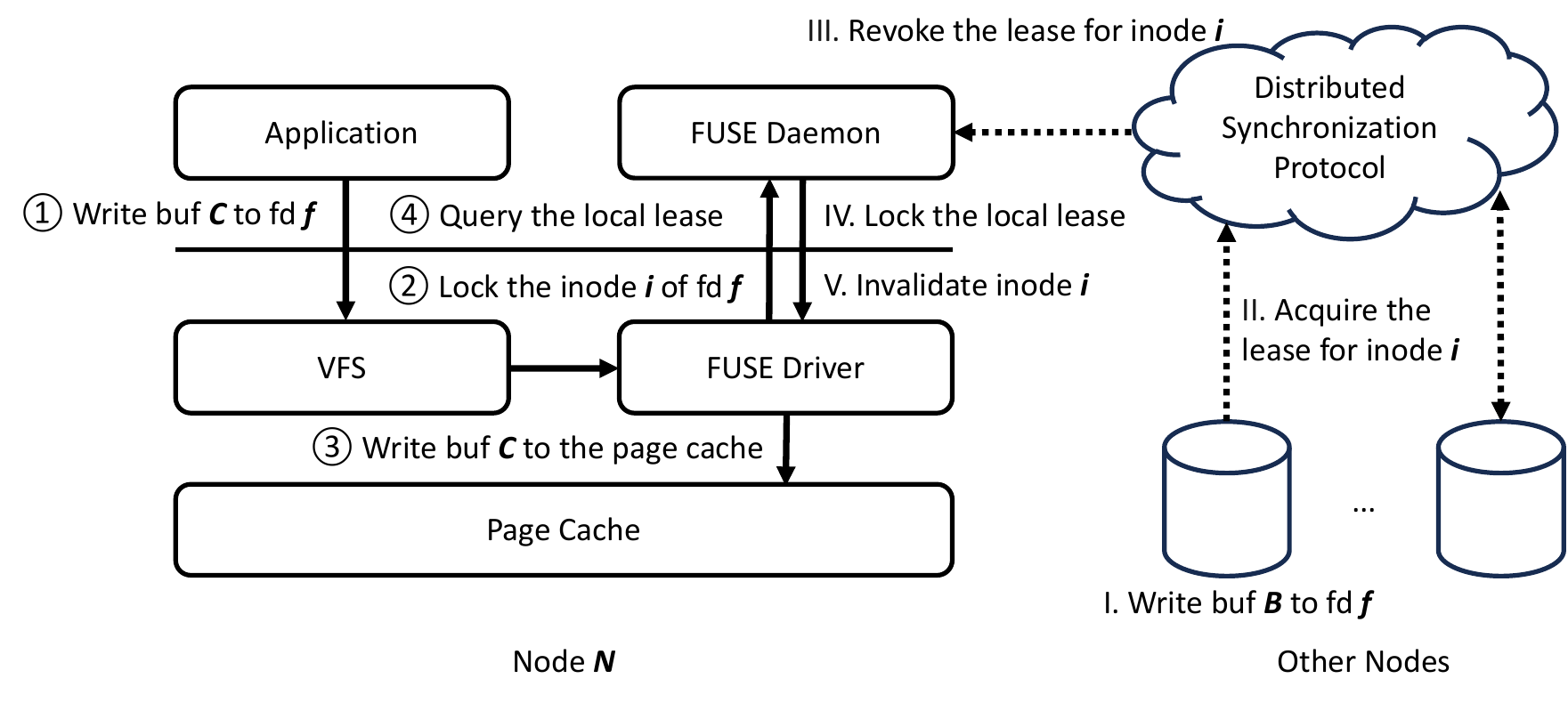}
        \caption{The deadlock issue caused by the reversed lock order during lease revocation in the write-through mode.}
        \label{fig:lock-eg}
    \end{subfigure}
    \caption{Challenges of leveraging the write-back page cache in FUSE-based distributed file systems}
    \label{fig:challenge}
\end{figure}

\subsection{Kernel-Userspace Coordination Failure in Write-Back Mode}\label{sec:challenge-consistency}
Write-back mode in the kernel page cache fundamentally breaks strong consistency in distributed FUSE systems. The primary reason is the architectural separation between where data operations occur (kernel) and where distributed coordination happens (userspace).

We illustrate the data inconsistency issue caused by enabling write-back cache in Figure~\ref{fig:write-back-eg}, where multiple nodes form a distributed file system and follow a synchronization protocol. To save space, we only plot the internal details of node \texttt{N}, since all nodes are symmetric. Suppose that node \texttt{N} writes some content \texttt{buf\ A} to file \texttt{f} (step \circled{1}). Under the write-back mode, the FUSE driver will only acquire the inode lock for \texttt{f} in the kernel (step \circled{2}) and then return a completion to the application after writing the data to the page cache (step \circled{3}). This data path bypasses the userspace daemon which implements the synchronization protocol between nodes. Consequently, not only are other nodes unable to read the updated content of \texttt{buf\ A} in file \texttt{f}, but also they themselves might write different values to the same file, leading to data inconsistency.

\subsection{Deadlock Caused by Reversed Lock Order With Write Through}\label{sec:challenge-lock}
As using the write-back mode bypasses any userspace synchronization, FUSE-based distributed file systems can resort to the write-through mode to utilize the kernel's page cache. As Figure~\ref{fig:write-through-eg} shows, after writing the data \texttt{buf\ A} to the page cache with the inode lock, to obtain knowledge of the distributed access permission, the FUSE driver issues a write-through operation to the userspace daemon (step \circled{4}). Then, the FUSE daemon is able to check its access permission for the file \texttt{f} (step \circled{5}) and send requests to acquire the write permission if necessary (step \circled{6}). So, write operations will only return to the application when the current node has the permission to do so, excluding data races shown in Figure~\ref{fig:write-back-eg}.

Nevertheless, there is an issue of locking order in the above write-through solution. Because different threads may acquire and revoke a lease concurrently, the file system usually needs to protect it with an additional read-write lock.

Figure~\ref{fig:lock-eg} presents an example. Suppose that node \texttt{N} has held a write lease for file \texttt{f} from the beginning. When writing \texttt{f}, it first needs to grab the inode lock in the kernel (step \circled{2}). Once the write request arrives in the userspace, it has to obtain another lock of the userspace lease to read its content (step \circled{4}). This kernel-userspace locking order would conflict with the sequence required during a lease revocation request. In particular, if another node wants to write the same file \texttt{f} (step \MakeUppercase{\romannumeral 1}) and tries to acquire the distributed lease of it (step \MakeUppercase{\romannumeral 2}), a revoke request for file \texttt{f}'s lease may be sent to node $N$ (step \MakeUppercase{\romannumeral 3}). At this time, the revoking request will grab locks in the reversed order: first locking the lease (step \MakeUppercase{\romannumeral 4}) and then the inode in the kernel (step \MakeUppercase{\romannumeral 5}), which leads to deadlocks. 

One workaround would be to use optimistic concurrency control (OCC). Specifically, at step \MakeUppercase{\romannumeral 4} of Figure~\ref{fig:lock-eg}, the revoker does not hold the lock of the lease when invalidating the inode, assuming that there is no concurrent writer that will update file $f$'s page cache. After the invalidation finishes, the revoker repeats the invalidation if any concurrent writers are detected. So, the deadlock issue is resolved. However, under high contention, the revoking performance would be low, leading to many aborts\cite{ye2023polaris, yuan2016bcc, huang2020opportunities}. Even worse, OCC would lead to unfair lease distribution, as the node that already has the lease can keep writing to the file even if it is being explicitly revoked\cite{ye2023polaris, bernstein1981concurrency}. 

In summary, it is challenging to leverage the kernel's page cache under the write-back or the write-through mode in a FUSE-based distributed file system. The fundamental reason is the lack of userspace information when writing to the page cache. Consequently, the file system either sacrifices the strong consistency (write back) or experiences high latencies of writes and revocations (write through). Systems like GlusterFS prioritize performance with write-back caching at consistency's expense, while others like Ceph-FUSE choose consistency through write-through implementations, accepting the performance penalty.

\section{Design}\label{sec:design}

\begin{figure}
    \centering
    \includegraphics[width=1\linewidth]{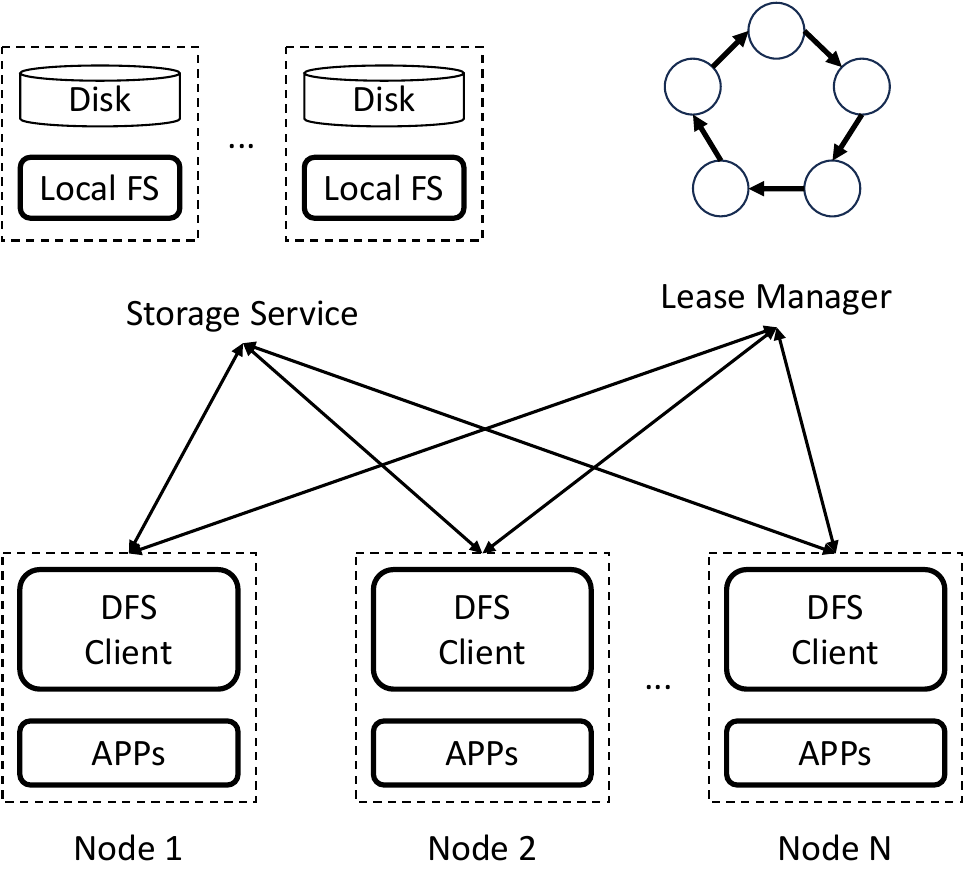}
    \caption{The overview of \name}
    \label{fig:dfuse-overview}
\end{figure}

\name is a FUSE-based distributed file system that enables write-back 
page caching while maintaining strong consistency. The system bridges the gap between high performance and strong consistency by offloading distributed concurrency control to the kernel driver, allowing the kernel's page cache to be safely utilized across distributed nodes. Table~\ref{tab:comparison} highlights \name's differences from existing FUSE-based and kernel-native solutions. Figure~\ref{fig:dfuse-overview} illustrates \name's overall architecture, which consists of three major components:

\begin{itemize}
    \item DFS client: Multiple userspace file system instances mounted on different nodes, providing POSIX-compliant file access to applications.
    \item Lease manager: A service that coordinates distributed leases to ensure consistency across nodes.
    \item Storage service: A centralized backend responsible for durable data storage.
\end{itemize}

These components run on different physical servers and communicate with each other via RPCs. To use \name, an application simply mounts a DFS client to a local directory (e.g., /mnt/fuse) and accesses files as if using a local file system. Applications across all nodes see a unified and consistent view of the file system through the standard POSIX interface.

\subsection{DFS Client}\label{sec:dfs}
\begin{figure}[t!]
    \centering
    \includegraphics[width=\linewidth]{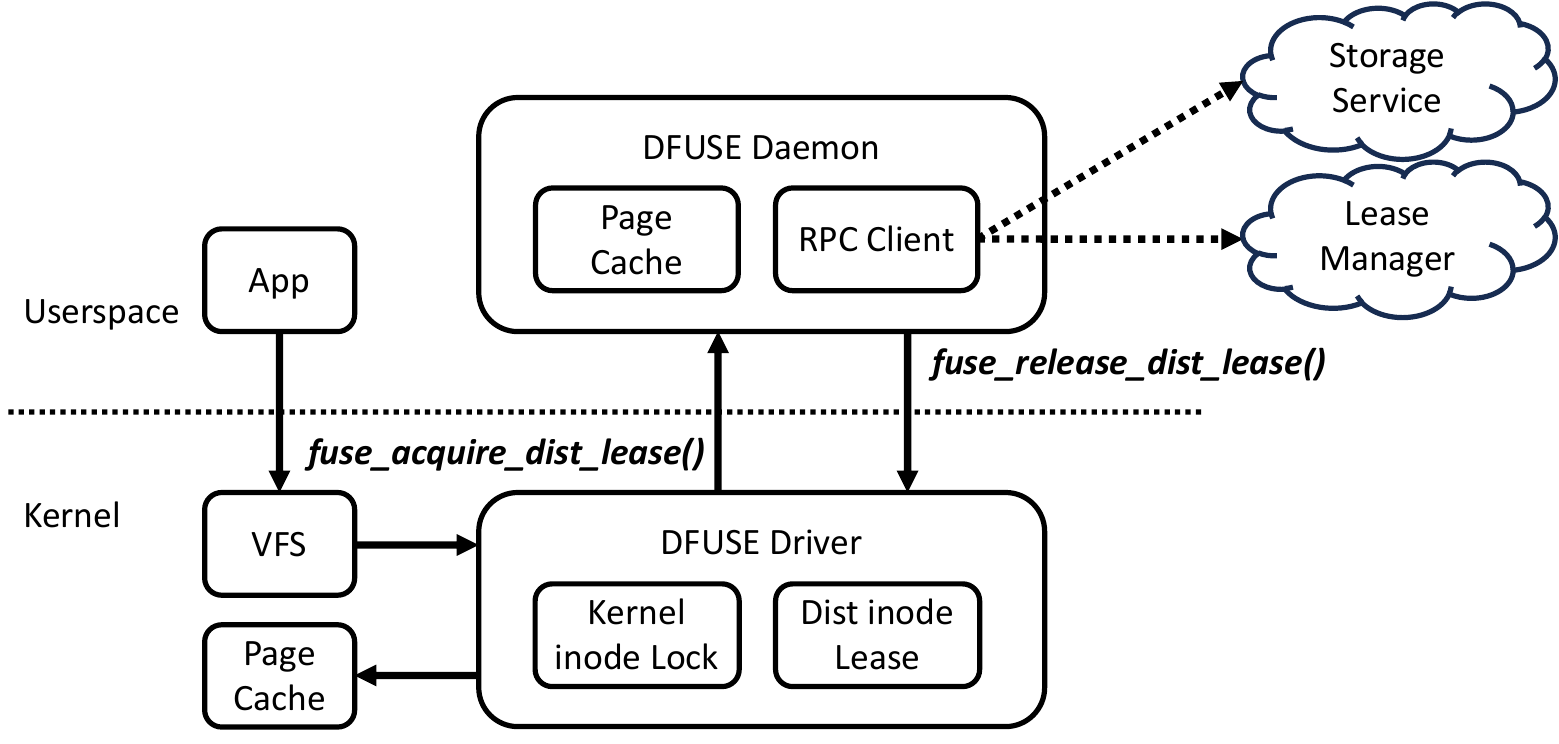}
    \caption{DFS client components.}
    \label{fig:dfs-instance}
\end{figure}

Each DFS client functions as a FUSE-based file system mounted to a local directory, collectively providing a unified namespace across all nodes. Applications interact with \name using standard POSIX interfaces. The key innovation in \name is offloading userspace distributed lease management to the kernel driver, enabling the system to validate lease status directly within the kernel during I/O operations.

This lease offloading mechanism is what enables \name to use write-back page caching while maintaining strong consistency across the distributed system. Figure~\ref{fig:dfs-instance} illustrates the internal components of a DFS client. Importantly, \name confines all kernel modifications to the FUSE kernel driver, leaving other kernel components (e.g., VFS and page cache) unchanged. Users simply need to load the \name kernel driver and install our customized libfuse.

\subsubsection{Why \name chooses leases}
Distributed file systems typically enforce cache consistency with callback locks~\cite{Howard1988AFS}, leases~\cite{gray1989leases}, version vectors~\cite{RFC3530}, or optimistic validation~\cite{Kung1981OCC}. We adopt leases because they already balance low-latency reads with bounded staleness and are widely deployed in production DFSs~\cite{Ghemawat2003GFS, Shvachko2010HDFS, Calder2011AzureStorage, Corbett2012Spanner, Ousterhout2012RAMCloud}. The lease protocol itself is therefore not a contribution of this work; its role is to provide a familiar yard-stick against which to evaluate \name's kernel-offload mechanism. \name's design also has the potential to utilize other consistency protocols mentioned above.

\subsubsection{Lease Offloading} 
\name's core innovation is embedding distributed lease information directly in the kernel's FUSE driver. In this section, we show how this design allows the system to coordinate page cache access across nodes without sacrificing performance. 

\name adds a distributed lease field to the FUSE inode data structure, representing the current node's access permission for that inode. This lease has three possible states: shared read, exclusive write, or null. The lease initializes to null when an application opens a file, indicating no initial read or write access.

For each inode operation requiring shared read or exclusive write, the \name driver checks the lease type after acquiring the kernel inode lock. If it finds that the node is not holding the appropriate lease (null or weaker than required), the driver requests the userspace daemon to update the lease. Conversely, when the \name daemon receives a revocation from the lease manager, it instructs the kernel driver to release the lease.

To manage these offloaded leases, \name introduces two new FUSE operations:

\paragraph{\texttt{fuse\_acquire\_dist\_lease()}} Called when the driver determines its local lease does not provide sufficient permission. The driver packages the inode number and desired permission (read or write) into a request sent to the userspace daemon through /dev/fuse. The daemon then acquires the distributed lease from the lease manager via RPC. Upon success, the driver updates the local lease and proceeds with the application's operation. The application blocks on a read or write until the FUSE driver finishes the operation.

While lease acquisition still requires an extra kernel-userspace round trip, subsequent accesses with the same lease can operate directly on the kernel page cache without userspace involvement, significantly reducing I/O latency.

\paragraph{\texttt{fuse\_release\_dist\_lease()}} Handles lease revocation by: 1) blocking new reads and writes, 2) waiting for ongoing operations to complete, 3) flushing dirty pages to storage, and 4) marking the local lease as null. This notification ensures the driver follows the same locking order as during lease acquisition, avoiding the deadlock issue described in \S\ref{sec:challenges}. If a revocation happens concurrently with application read or write requests (to the same file), the application will have to wait until the FUSE driver re-acquires the lease.

\paragraph{Example end-to-end workflow.} To better demonstrate the lease offloading mechanism in \name, we show the workflow of a write request from an application running on DFS client \texttt{A}. First, the application issues a write syscall with glibc, which will be routed to the FUSE kernel driver by VFS. Next, the FUSE driver locks the lease of the inode (serialize concurrent revocation) and checks the lease status. If the lease has the write permission, then the FUSE driver proceeds with the normal write-back operation, i.e., acquiring the inode lock and updating the page cache. However, if the lease status is either null or shared read, the FUSE driver will send a \texttt{fuse\_acquire\_dist\_lease()} operation to the userspace daemon through /dev/fuse. The kernel thread falls into interruptible sleep while waiting for the result. Once the userspace daemon acquires a write lease from the manager, it wakes up the FUSE driver thread. Lastly, the driver updates the lease type to write, updates the kernel page cache, and returns the control to the application. This solves the challenge mentioned in \S\ref{sec:challenge-consistency}.

We next discuss how a DFS client handles lease revocations. Suppose that during a write request, the userspace daemon of client \texttt{A} receives a lease revocation request for the same file. To properly revoke the lease, client \texttt{A} must avoid new I/Os to this file and the deadlock issue mentioned in \S\ref{sec:challenge-lock}. So, the daemon triggers a \texttt{fuse\_release\_dist\_lease()} notification to the FUSE driver, which will first lock the lease (to prevent new I/Os from reading the status) and then lock the inode (to flush its cached pages). By following the same locking order as normal read and write requests, \name resolves the deadlock challenge in \S\ref{sec:challenge-lock}. Subsequent I/Os to the file will find the lease in the null status and trigger the above lease acquisition.

\subsubsection{Userspace Buffer Cache}
To further optimize performance and reduce communication with the remote storage service, \name implements a secondary in-memory buffer cache in the userspace daemon. This inclusive userspace cache prevents network bottlenecks by batching operations that would otherwise require immediate network round trips when data is flushed from kernel, and it enables customized caching policies in the userspace. Unlike the kernel's page cache, which can grow dynamically based on system memory pressure, the userspace buffer cache maintains a fixed memory reservation to provide predictable performance and resource utilization.

This dual-caching strategy creates a tiered approach to data access. When the kernel's background threads asynchronously flush dirty pages from the kernel page cache, the data is not immediately sent to the remote storage service. Instead, these pages are first buffered in the userspace buffer cache, which serves as an intermediary layer between the kernel and remote storage.

Similarly, when a read operation encounters a cache miss in the kernel page cache, the \name driver forwards the request to the userspace daemon. The daemon first attempts to locate the requested data in its userspace buffer cache before making any RPC calls to the remote storage service. This approach effectively reduces the frequency of expensive remote storage operations, especially for workloads with temporal locality beyond what the kernel cache can capture.

To further minimize RPC overhead, \name implements batch read and write operations that consolidate multiple page accesses into a single RPC request. When the userspace daemon needs to read or write multiple pages, it aggregates these operations and sends them as a batch to the remote storage service, amortizing the cost of network latency and protocol overhead across multiple operations. This batching mechanism is particularly beneficial for sequential I/O patterns and large file transfers where multiple adjacent pages are likely to be accessed together.

As we will show in the evaluation (\S\ref{sec:eval}), the combination of kernel page caching, userspace page caching, and batch operations significantly reduce the performance impact of remote storage access while maintaining strong consistency guarantees across the distributed system.

\subsubsection{Global File Identifier}
Since FUSE reuses existing data structures from VFS, its inode number generation is only based on local assignment. Therefore, every DFS client may have different inode numbers for the same file. Fortunately, FUSE provides a tag for each file to save custom fields and passes the tag in FUSE operations. \name stores a global file identifier (GFI) in the tag to identify the same files among different DFS clients. Specifically, the GFI consists of two parts: 1) the ID of the node in the storage service that actually stores the file and 2) the local inode number from that storage node. Both DFS clients and the lease manager use GFIs to identify shared files. For example, any messages of acquiring and revoking leases contain the GFI of the target file, and a DFS client can determine the target storage node when flushing dirty pages of a file.

\subsection{Lease Manager}\label{sec:lease}
\name employs a lease-based design to coordinate distributed file access. The system maintains a conceptually global lease manager that enforces strong consistency through per-file distributed read-write leases. This lease mechanism follows the classic distributed read-write lock pattern, with algorithms detailed in Algorithm~\ref{alg:algo-dfs} and Algorithm~\ref{alg:algo-mgr} for DFS clients and the lease manager, respectively.

To avoid a single point of failure, the lease manager can be replicated either by running a consensus protocol such as Paxos~\cite{lamport2001paxos} or Raft~\cite{ongaro2014search}, or by outsourcing the state to an off-the-shelf coordination service such as ZooKeeper~\cite{hunt2010zookeeper}. DFS clients communicate with the lease manager via RPCs. If the lease manager becomes unreachable due to network partition, clients retain the lease until its soft timeout and fall back to write-through on lease expiry, preserving correctness. At any time, a file may have at most one exclusive writer or multiple readers. Before each read or write operation, a DFS client calls AcquireLease (Algo~\ref{alg:algo-dfs}) to ensure it holds the appropriate lease for the target inode.

If the local lease is invalid (e.g., expired) or insufficient for the intended operation (e.g., attempting to write with only a read lease), the DFS client requests an updated lease from the lease manager. A DFS client may also receive revocation requests from the manager explicitly requesting release of a lease. In this case, the instance clears cached data for the inode and flushes any dirty pages back to the storage service, ensuring other nodes can access the latest content.

The lease manager maintains information about current lease types and owners for each file. When receiving lease applications, it determines the appropriate action based on the global lease state and the requesting node's intent. For compatible requests (e.g., multiple read leases), the manager simply adds the requesting node to the owner list. For conflicting requests (e.g., when either the current owner or requester needs write access), the manager revokes existing leases before granting the new lease.

\begin{algorithm}[t!]
\caption{Lease Algorithm on DFS clients}\label{alg:algo-dfs}
\begin{algorithmic}[1]
\Procedure{AcquireLease}{$inode$, $intent$}
\State $lease \gets \Call{GetLocalLease}{inode}$
\If {$lease.type$ satisfies $intent$}
    \State \Return
\EndIf
\If{$lease.type = READ \land intent = WRITE$}
    \State \Call{ReleaseLease}{$inode$}
    \State $manager$.\Call{RemoveOwner}{$inode$, $self$}
\EndIf
\State $manager$.\Call{GrantLease}{$inode$, $intent$, $self$}
\State $lease.type \gets intent$
\EndProcedure

\Procedure{ReleaseLease}{$inode$}
\State Clear the page cache of $inode$
\State Sync dirty pages to the storage service
\State $inode \gets None$
\EndProcedure
\end{algorithmic}
\end{algorithm}

\begin{algorithm}[t!]
\caption{Lease Algorithm on Lease Manager}\label{alg:algo-mgr}
\begin{algorithmic}[1]
\Procedure{GrantLease}{$inode$, $intent$, $node$}
\State $lease \gets \Call{GetLocalLease}{inode}$
\If {$lease.owner \ is \ empty$}
    \State $lease.type \gets intent$
    \State $lease.owner \gets node$
\ElsIf {$lease.type = READ \land intent = READ$}
    \State Add $node$ to $lease.owner$
\ElsIf{$lease.type = READ \land intent = WRITE$}
    \For{$holder \in lease.owner$}
        \State $holder$.\Call{ReleaseLease}{$inode$}
    \EndFor
    \State $lease.type \gets WRITE$
    \State $lease.owner \gets node$
\ElsIf{$lease.type = WRITE$}
    \State $holder \gets lease.owner$
    \State $holder$.\Call{ReleaseLease}{$inode$}
    \State $lease.type \gets intent$
    \State $lease.owner \gets node$
\EndIf
\EndProcedure
\end{algorithmic}
\end{algorithm}

\subsection{Storage Service}
\name separates data storage from file system processing by implementing a dedicated storage service. This design aligns with modern cloud computing's disaggregation of compute and storage resources. Rather than reinventing storage management, \name leverages existing mature storage services (such as S3\cite{noauthor_amazon_nodate}, GCS\cite{noauthor_cloud_nodate-1}, or Pangu\cite{li2023more}) that already provide high reliability and availability. These services offer block or object storage interfaces sufficient to support POSIX file system semantics.

This separation simplifies DFS clients, as they no longer need to manage storage resource allocation or data replication. The design also acknowledges that local storage attached to cloud VMs (e.g., Amazon's ECS) is typically ephemeral—storing data locally on client VMs would risk data loss upon VM failures.

In environments where cloud storage services are unavailable or undesirable, such as on-premise clusters, the storage service can alternatively be implemented as a replicated storage cluster. This approach still offers benefits like more efficient resource allocation. The number and implementation of storage nodes remain decoupled from DFS clients, allowing flexibility to select storage backends (e.g., SPDK\cite{yang2017spdk}, RAID\cite{chen1994raid}, Ext4\cite{mathur2007new}, XFS\cite{sweeney1996scalability}) based on workload characteristics.

\subsection{Adoption by Existing Systems}
\name's architecture is designed with portability and extensibility in mind, making it readily adaptable for existing FUSE-based distributed file systems. The core innovation of lease offloading is implemented as a modular extension to the standard FUSE kernel driver. Existing systems like GlusterFS, Ceph-FUSE, or JuiceFS can adopt \name's lease offloading mechanism with three straightforward integration steps: 1) replace their standard FUSE driver with \name's modified kernel driver, 2) adapt their userspace daemon to implement the two new FUSE operations (\texttt{fuse\_acquire\_dist\_lease()} and \texttt{fuse\_release\_dist\_lease()}), and 3) integrate with a lease manager service or extend their existing metadata coordination services to handle lease management. This approach allows these systems to maintain their established storage backends, authentication mechanisms, and specialized optimizations while gaining \name's ability to leverage write-back caching with strong consistency guarantees.
\section{Implementation}\label{sec:impl}
We implemented \name on libfuse 3.16 and Linux 5.10. We selected gRPC as the inter-node communication protocol. For this prototype implementation, the storage service is comprised of multiple ext4 file systems, where each distributed file system instance translates low-level FUSE operations to their corresponding ext4 operations. To manage the userspace buffer cache efficiently, we integrated Facebook's CacheLib framework\cite{berg2020cachelib}, which enables flexible admission and eviction policies with minimal overhead. The implementation required approximately 4,000 lines of code for the userspace file system component, with relatively minimal modifications to the existing codebase: approximately 100 lines of code for libfuse modifications and 300 lines for the FUSE kernel driver enhancements. 
\section{Evaluation}\label{sec:eval}
We seek to answer three evaluation questions:

\textbf{Q1:} How does \name perform end-to-end under different workloads (\S\ref{sec:eval-workloads})?

\textbf{Q2:} How do various contention levels affect \name's performance (\S\ref{sec:eval-contention})?

\textbf{Q3:} How does \name scale with increasing number of nodes (\S\ref{sec:eval-scalability})?

\begin{figure*}[t!]
    \centering
    \begin{subfigure}{0.45\textwidth}
        \centering
        \includegraphics[width=1\linewidth]{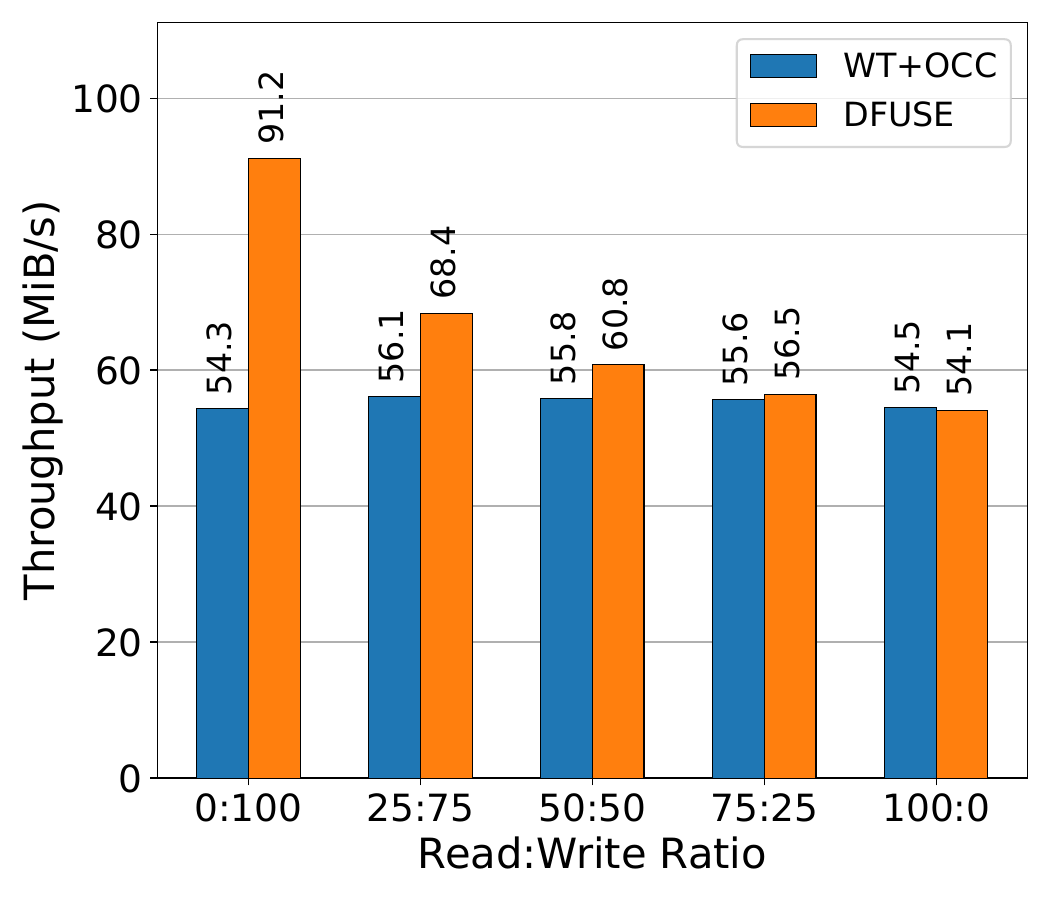}
        \caption{fio Random I/O (Throughput)}
        \label{fig:fio-rand-thru}
    \end{subfigure}
    \hfill
    \begin{subfigure}{0.45\textwidth}
        \centering
        \includegraphics[width=1\linewidth]{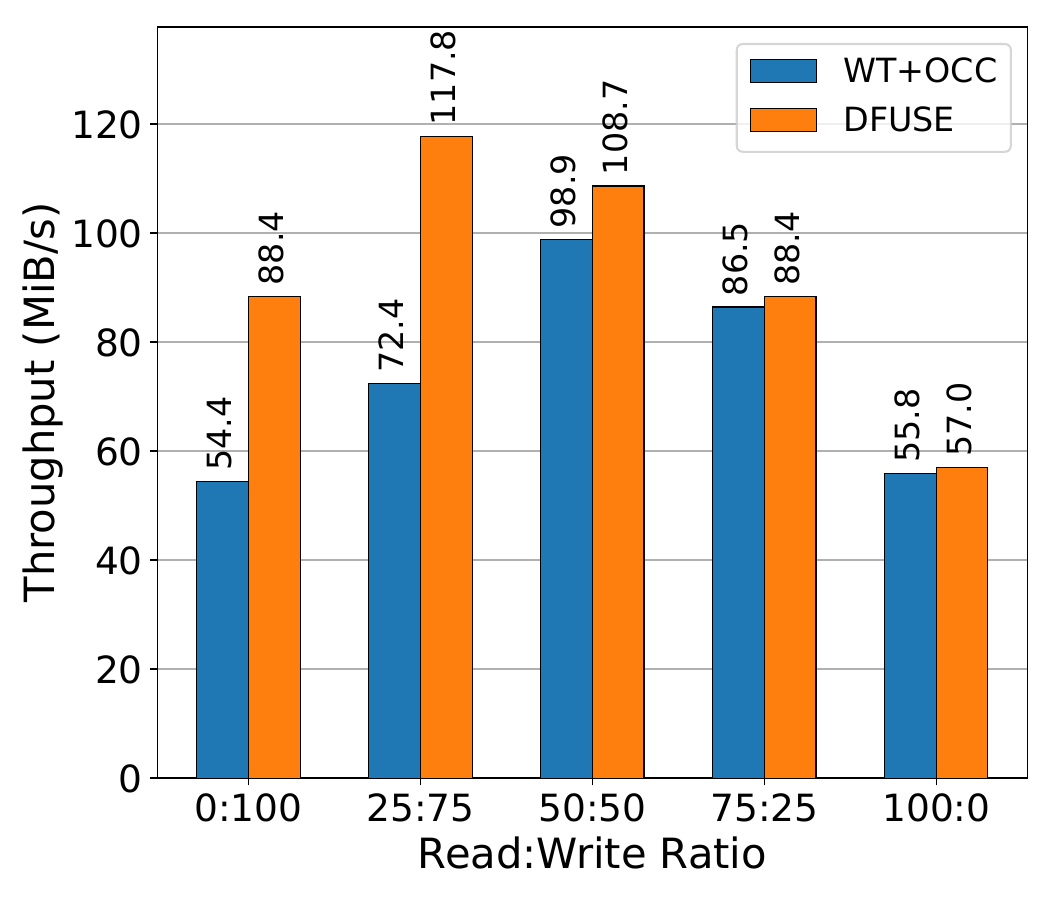}
        \caption{fio Sequential I/O (Throughput)}
        \label{fig:fio-seq-thru}
    \end{subfigure}
    \hfill
    \begin{subfigure}{0.45\textwidth}
        \centering
        \includegraphics[width=1\linewidth]{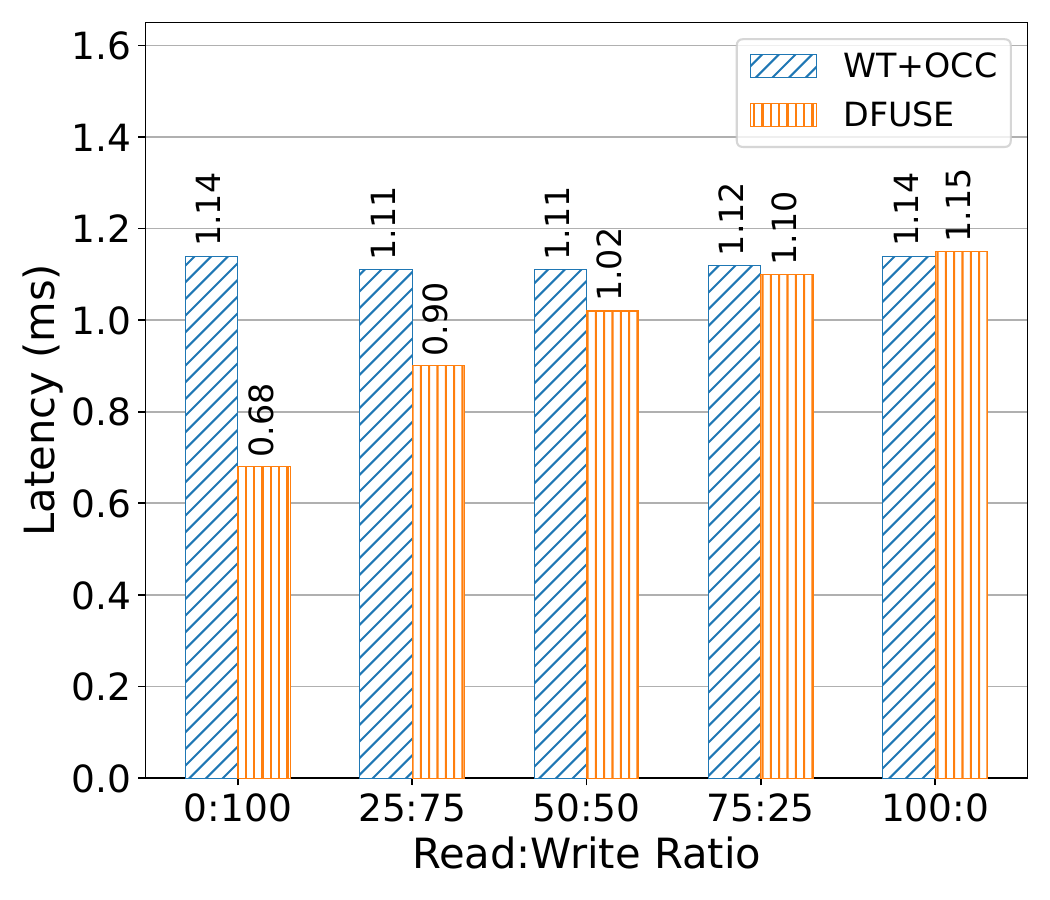}
        \caption{fio Sequential I/O (Latency)}
        \label{fig:fio-rand-lat}
    \end{subfigure}
    \hfill
    \begin{subfigure}{0.45\textwidth}
        \centering
        \includegraphics[width=1\linewidth]{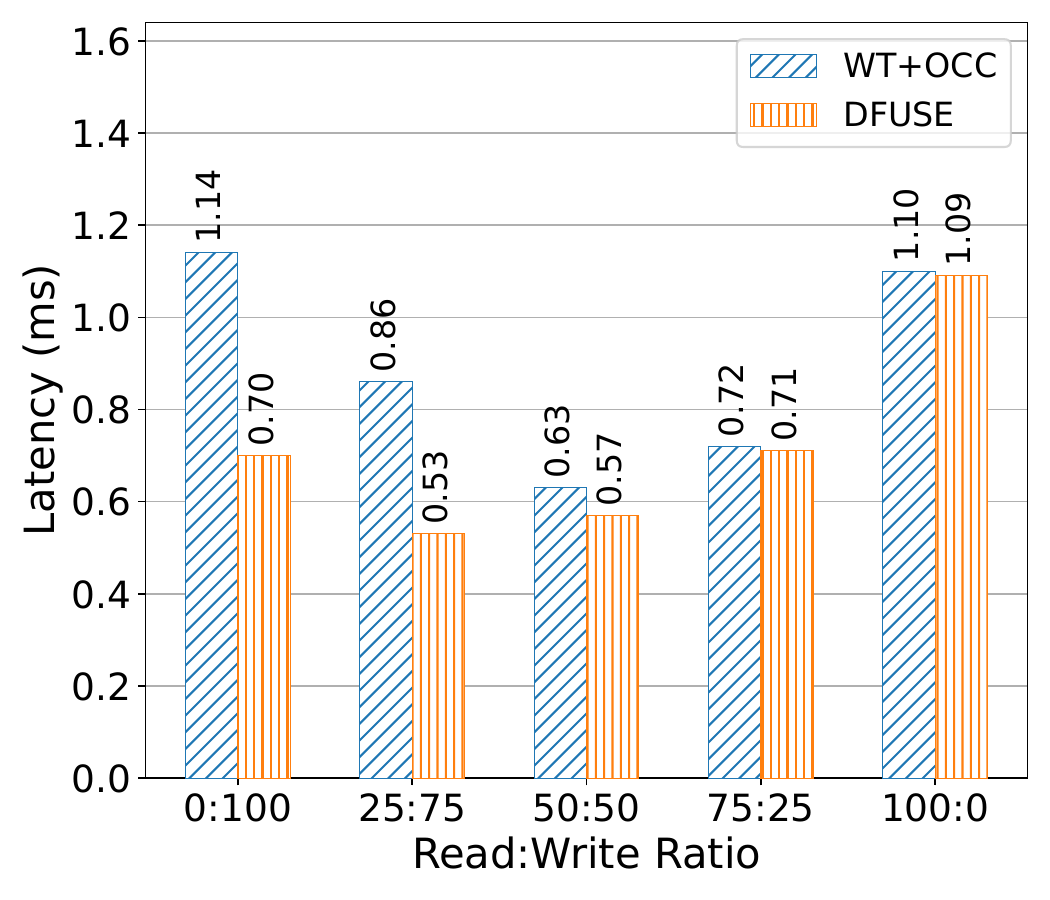}
        \caption{fio Sequential I/O (Latency)}
        \label{fig:fio-seq-lat}
    \end{subfigure}
    \hfill
    \caption{\name's throughput and latency under different fio workloads without contention}
    \label{fig:fio}
\end{figure*}

\subsection{Experiment Setup}

\begin{table}[tb]
    \centering
    \caption{Filebench workload characteristics}
    \label{tab:filebench-characteristics}
    \begin{tabular}{lrrrc}
        \hline
        \textbf{Workload} & \textbf{\# Files} & \textbf{Dir Width} & \textbf{File Size} & \textbf{R/W Ratio} \\
        \hline
        Fileserver  & 10,000    & 20         & 1.25MB   & 1:2    \\
        Webserver   & 80,000    & 20         & 160KB    & 10:1   \\
        Netsfs      & 74,000    & 20         & 267KB    & 5:2    \\
        \hline
    \end{tabular}
\end{table}

\paragraph{Setup.}  We conduct all experiments on CloudLab's \cite{ricci2014introducing, duplyakin2019design} c220g1 servers. Each c220g1 server has two 8-core Intel E5-2630 CPUs at 2.4 GHz, 128 GB DDR4 memory, a 480 GB Intel DC S3500 SSD, and a dual-port Intel X520-DA2 10Gb NIC~\footnote{The second port was unused in the experiment.}. The OS is Ubuntu 20.04 LTS with a Linux version of 5.10. 

In our evaluation, unless otherwise specified, we run \name with four DFS clients, one storage node with ext4 file system, and one lease manager node. For each DFS client, we mount the FUSE daemon with default permissions and multi-threading. Both the baseline and \name use a userspace cache of 1 GB per instance.

\paragraph{Workload.} We use fio~\cite{axboe_flexible_2022} and filebench~\cite{noauthor_filebenchfilebench_2025} as micro and macro benchmarks, respectively, to evaluate \name. In fio, the patterns cover both random and sequential I/O with five read-write ratios. There are four threads per node, each operating on a working set of 100 16MB files. We choose such a working set size to emulate SSTable objects (typically sized in the tens of MB)~\cite{facebook_compaction_nodate, noauthor_tune_nodate}, generate enough hot inodes to exercise leases/invalidations, and avoid saturating storage/network--our conclusions target per-inode coherence costs rather than absolute dataset size. We run three representative applications from filebench in this evaluation: fileserver, webserver, and netsfs (their characteristics are shown in Table~\ref{tab:filebench-characteristics}).

\paragraph{Metric.} We compare the aggregated throughput and average latency to reflect \name's performance. The throughput reported in this section is the cluster-wide result, summing each node's throughput. The latency is the arithmetic mean of all operations' latency.

\paragraph{Baseline.} Our goal is to isolate the effect of kernel write-back vs. write-through under identical file system logic, so we compare against a same-code baseline that differs only in cache/write policy (WT+OCC). As we discussed in \S\ref{sec:challenges}, this is the best that current systems like Ceph-FUSE\cite{weil2006ceph} and GlusterFS\cite{noauthor_gluster_nodate} can do if they wish to leverage kernel's page cache while ensure strong consistency. Head-to-head with Ceph/Gluster would conflate client, metadata, striping, recovery, and placement policies with our lease offloading mechanism, making causal attribution unclear.

\subsection{Different Workloads (Q1)}\label{sec:eval-workloads}
To evaluate \name's end-to-end performance across various workloads, we conducted experiments using both random and sequential I/O patterns with varying read-write ratios. Figures~\ref{fig:fio} present the total throughput and average latency comparison between \name and the baseline (write-through with OCC) implementation.
For random I/O workloads (Figure~\ref{fig:fio-rand-thru} and ~\ref{fig:fio-rand-lat}), \name demonstrates consistent performance advantages across all workload with writes. The improvement is particularly pronounced in write-heavy scenarios (i.e., 0:100 and 25:75 read-write ratios), where \name achieves up to 68.0\% higher throughput and 40.4\% lower latency. This significant improvement stems from \name's ability to leverage write-back caching without synchronizing every write to the userspace. As the workload becomes more read-intensive, the performance gap narrows but remains favorable for \name, with improvements of 9.0\% for balanced workloads (50:50) and 1.6\% for read-heavy workloads (75:25). For pure read workloads (100:0), both systems perform similarly since read operations benefit equally from the kernel's page cache under both write-back and write-through mode.

Sequential I/O results (Figure \ref{fig:fio-seq-thru} and \ref{fig:fio-seq-lat}) show even more substantial improvements. \name outperforms the baseline by 62.2\% for write-only workloads (0:100) and 62.7\% for mostly-write workloads (25:75). This larger improvement in sequential patterns can be attributed to \name's efficient batching of operations and the tiered caching approach that reduces kernel-userspace transitions. Even for balanced (50:50) and read-dominant workloads (75:25), \name maintains improvements of 9.9\% and 2.2\% respectively. The lack of improvement for pure read workloads (100:0) is expected, as both systems can effectively utilize the kernel's page cache for reads. The latency results in Figure \ref{fig:fio-seq-lat} follow the same trend.

The application-level benchmarks (Figure \ref{fig:filebench}) further validate \name's performance advantages across realistic workloads. In the fileserver benchmark, which simulates a write-heavy file server workload, \name achieves 26.2\% higher throughput and 20.0\% lower latency under no contention. The Netsfs benchmark shows substantial throughput gains of 18.5\% (under no contention), too. The webserver benchmark, which is predominantly read-oriented, shows a decrease of 11.0\% under no contention, suggesting some optimization opportunities for specific read-dominated patterns.

\begin{figure}[t!]
    \centering
    \includegraphics[width=\linewidth]{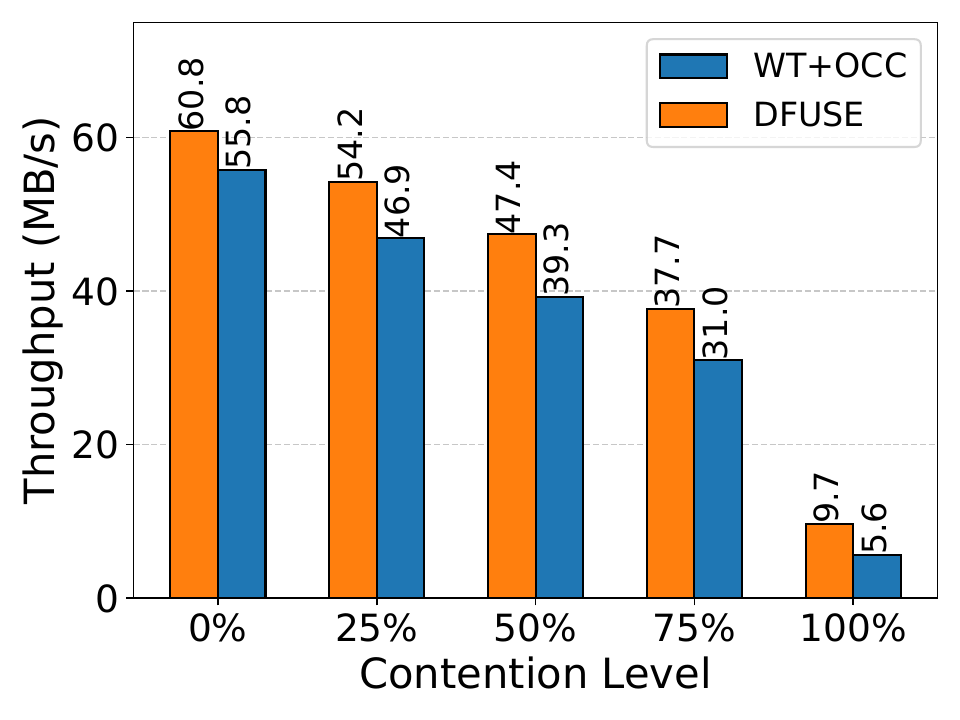}
    \caption{\name under various contention levels (with a 50:50 random read-write fio workload)}
    \label{fig:fio-cl}
\end{figure}

\subsection{Contention Levels (Q2)}\label{sec:eval-contention}
\paragraph{Definition.} A critical aspect of distributed file systems is performance under varying levels of contention. We define the contention level as the percentage of shared files with other nodes in one's working set. We treat any concurrent accesses (read or write) on the same file as contention, because even multiple readers will also trigger the distributed consistency mechanism, i.e., acquiring a shared read lease.

\paragraph{Setup.} Following the above definition, each node divides its working set into shared and private files. All nodes in the cluster will operate on the shared files concurrently, whereas the private files are only accessed by one node. For example, a contention level of 25\% in Figure \ref{fig:fio-cl} means that each node reads/writes 25 globally shared files and operates on 75 private files without other contenders (a working set of 100 files). We control the contention level by first creating all files on the storage nodes before the experiment starts, then specifying the working set of each node by file names during runtime.\\

Figure \ref{fig:fio-cl} illustrates that \name constantly outperforms the baseline as contention increases from 0\% to 100\%. At low contention (0\%), \name delivers 60.8 MB/s throughput compared to the baseline's 55.8 MB/s, representing an 9.0\% performance advantage. This benefit stems from \name's efficient use of write-back caching without the overhead of frequent lease revocations. As contention increases to moderate levels (25-50\%), both systems experience performance degradation, though \name maintains its edge with 54.2 MB/s at 25\% contention (15.6\% higher than baseline's 46.9 MB/s) and 47.4 MB/s at 50\% contention (20.6\% higher than baseline's 39.3 MB/s). This trend demonstrates \name's ability to effectively manage lease revocations while minimizing their impact on performance.

Under high contention (75-100\%), the absolute performance of both systems decreases substantially, but \name's relative advantage grows. At 75\% contention, \name delivers 37.7 MB/s versus the baseline's 31.0 MB/s (21.6\% higher). At extreme contention (100\%), \name still achieves 9.7 MB/s compared to the baseline's 5.6 MB/s, representing a 73.2\% advantage. This demonstrates that \name's kernel-level lease management approach significantly reduces the overhead of maintaining strong consistency even under challenging workload conditions.

\begin{figure*}[t!]
    \centering
    \begin{subfigure}[b]{0.3\textwidth}
        \centering
        \includegraphics[width=\textwidth]{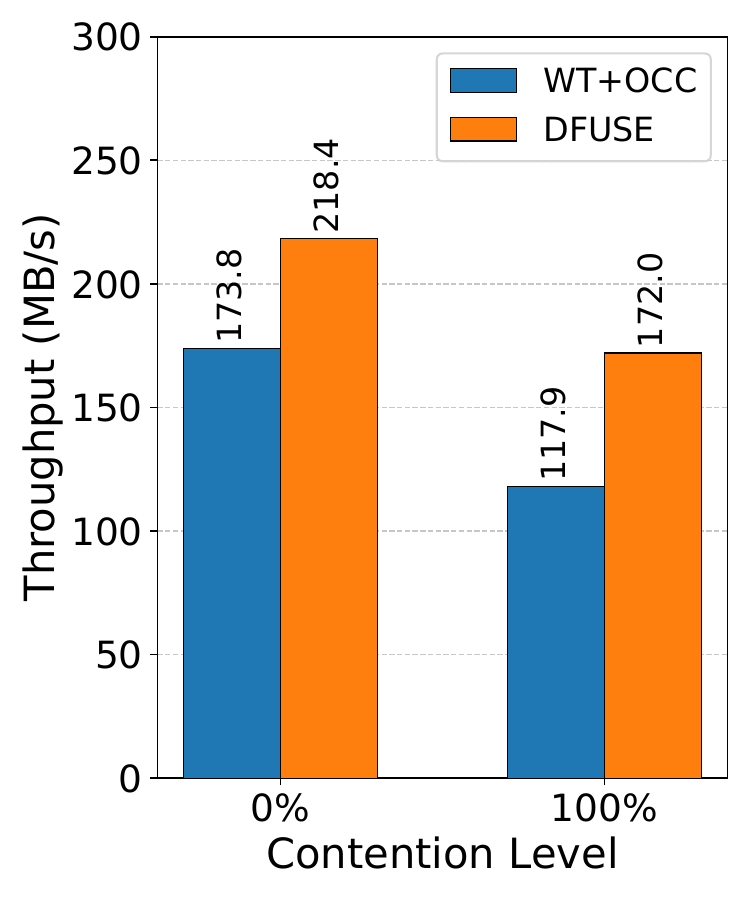}
        \caption{Fileserver (Throughput)}
        \label{fig:fileserver}
    \end{subfigure}
    \hfill
    \begin{subfigure}[b]{0.3\textwidth}
        \centering
        \includegraphics[width=\textwidth]{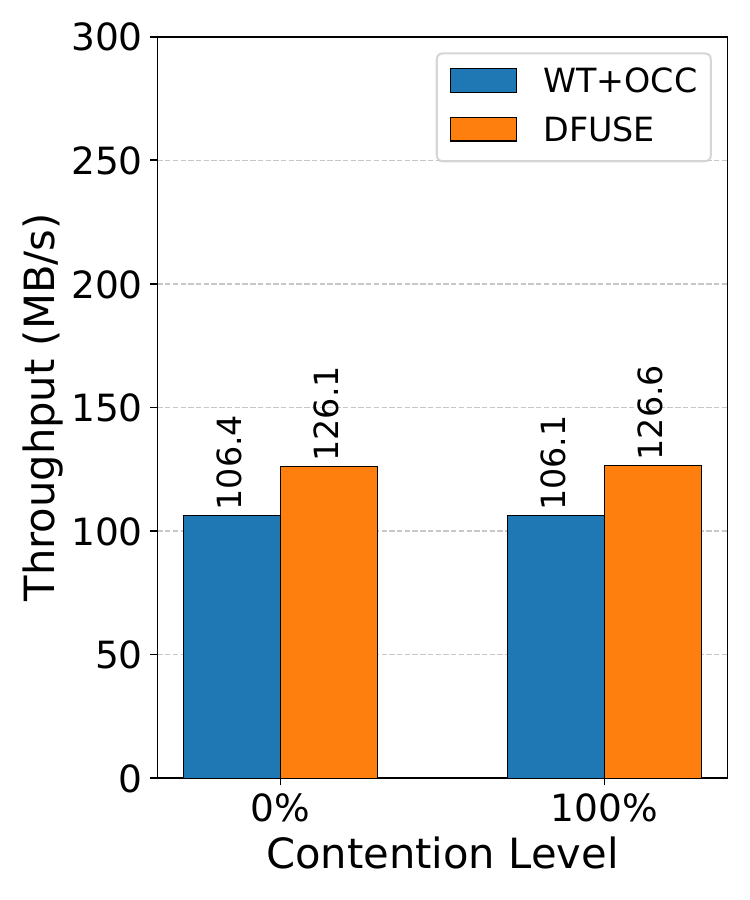}
        \caption{Netsfs (Throughput)}
        \label{fig:nfs}
    \end{subfigure}
    \hfill
    \begin{subfigure}[b]{0.3\textwidth}
        \centering
        \includegraphics[width=\textwidth]{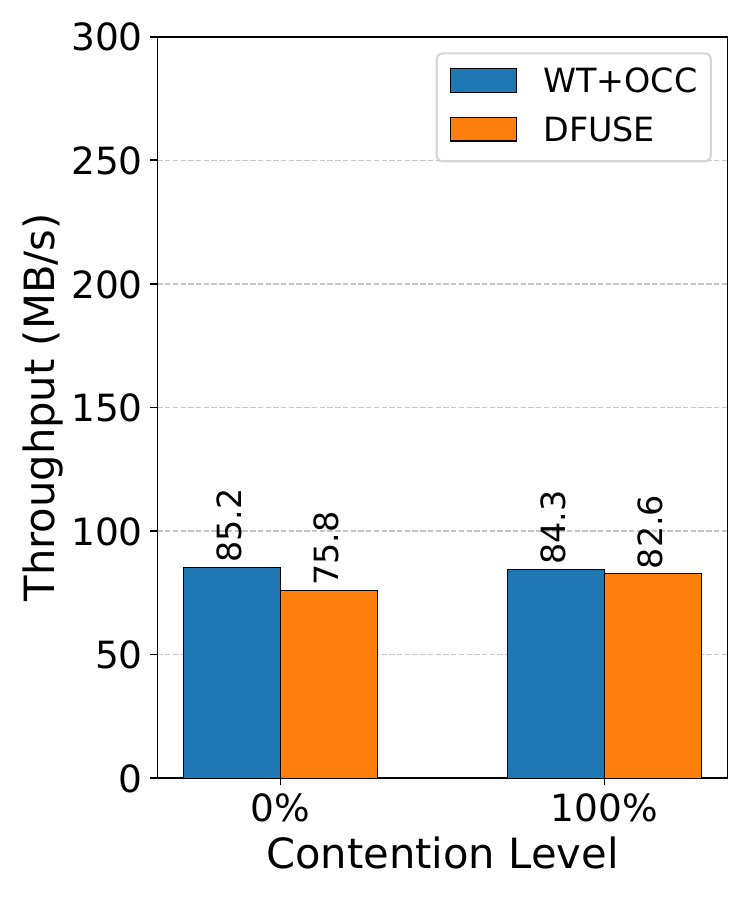}
        \caption{Webserver (Throughput)}
        \label{fig:webserver}
    \end{subfigure}
    \begin{subfigure}[b]{0.3\textwidth}
        \centering
        \includegraphics[width=\textwidth]{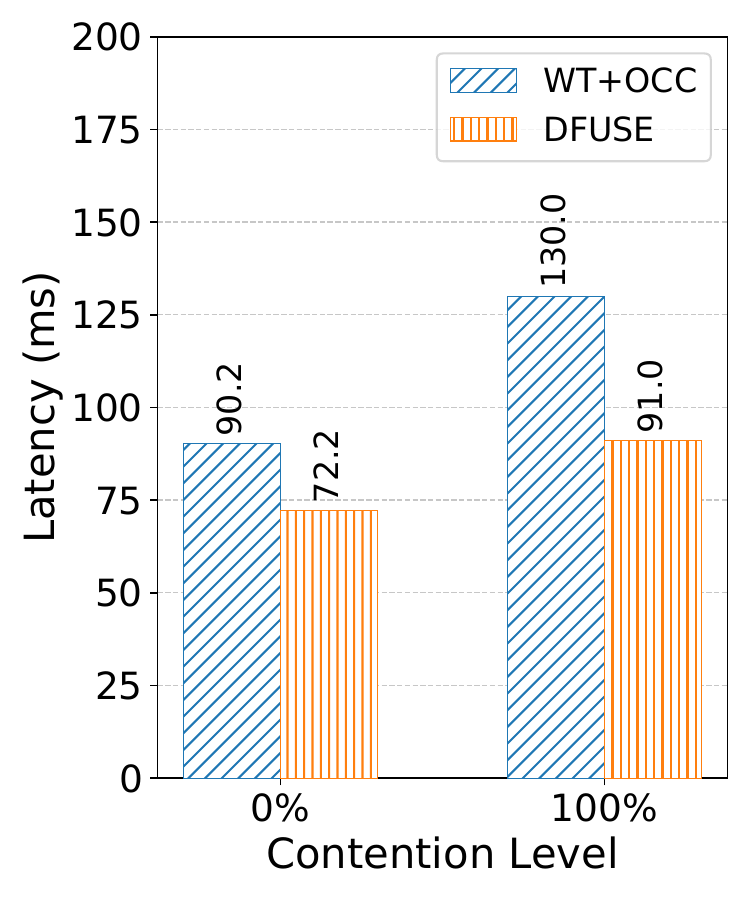}
        \caption{Fileserver (Latency)}
        \label{fig:fileserver-lat}
    \end{subfigure}
    \hfill
    \begin{subfigure}[b]{0.3\textwidth}
        \centering
        \includegraphics[width=\textwidth]{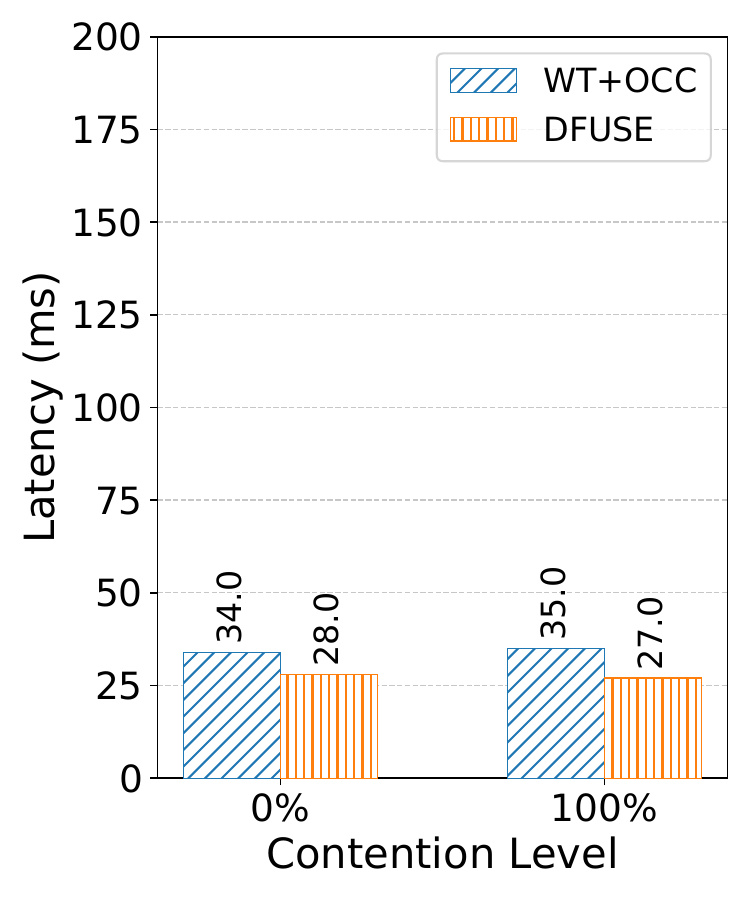}
        \caption{Netsfs (Latency)}
        \label{fig:nfs-lat}
    \end{subfigure}
    \hfill
    \begin{subfigure}[b]{0.3\textwidth}
        \centering
        \includegraphics[width=\textwidth]{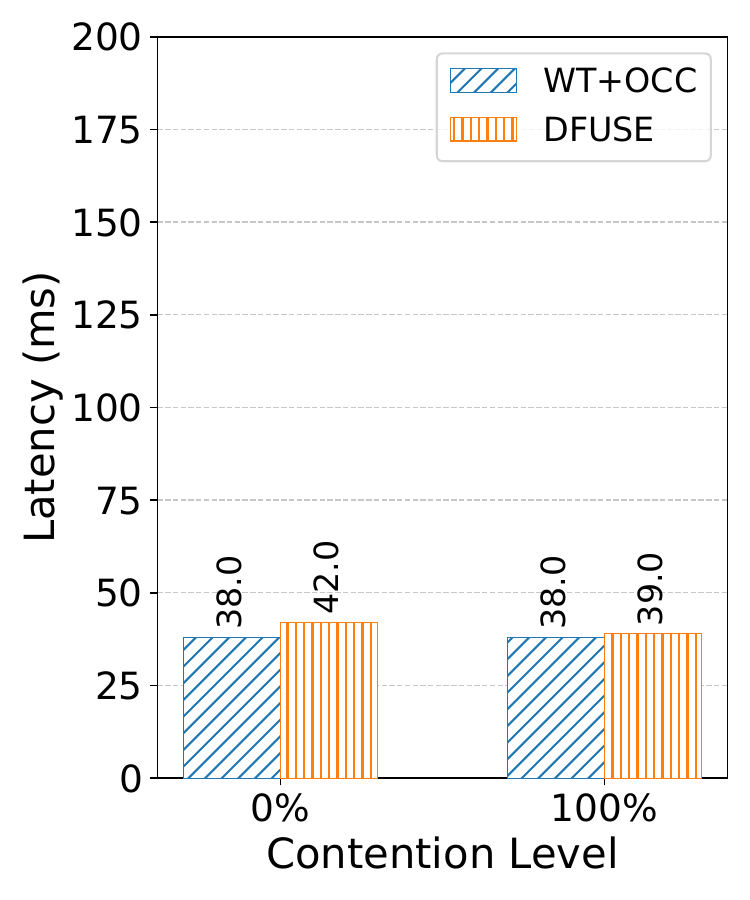}
        \caption{Webserver (Latency)}
        \label{fig:webserver-lat}
    \end{subfigure}
    \caption{\name under various contention levels (filebench applications)}
    \label{fig:filebench}
\end{figure*}

The application benchmarks in Figure \ref{fig:filebench} provide further insight into real-world performance under contention. For the Fileserver workload, \name achieves 221.2 MB/s with no contention and maintains 171.4 MB/s under contention (a 22.5\% reduction), whereas the baseline drops from 175.3 MB/s to 125.8 MB/s (a 28.2\% reduction). For the read-heavy Netsfs and Webserver workloads, changing contention from 0\% to 100\% hardly affects either design because both applications are dominated by reads and thus trigger few lease revocations.

These results confirm that \name's kernel offloading mechanism effectively balances strong consistency with high performance across varying contention levels. By embedding lease information directly in the kernel driver, \name minimizes the performance impact of lease management and enables efficient write-back caching even in high-contention scenarios.

\begin{figure}[t!]
    \centering
    \includegraphics[width=\linewidth]{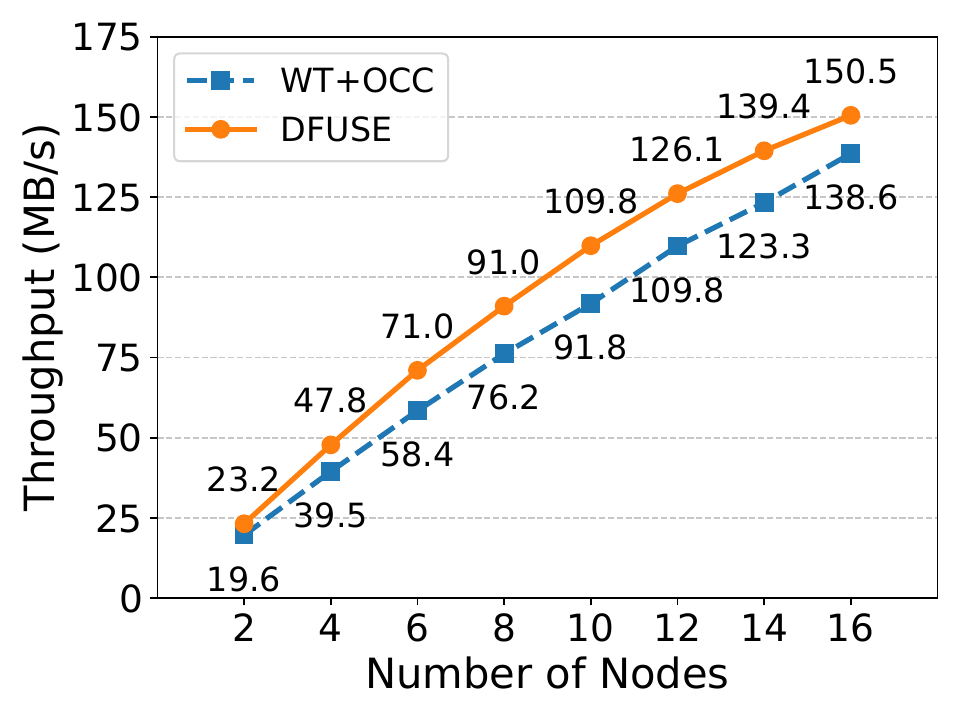}
    \caption{\name scales with increasing number of nodes (with a contention level of 50\% and a 50\%-50\% random read-write workload)}
    \label{fig:fio-scalability}
\end{figure}

\subsection{Scalability (Q3)}\label{sec:eval-scalability}
Figure \ref{fig:fio-scalability} illustrates \name's scalability as the number of nodes (i.e., the count of client machines) increases from 2 to 16. The workload is 50\%-50\% random read-write with fio, and we set the contention level to 50\%. Both \name and the baseline demonstrate near-linear scaling, with throughput increasing proportionally to the number of nodes, suggesting neither system suffers from significant coordination bottlenecks.

At smaller scales (2-4 nodes), \name demonstrates a significant performance advantage over the baseline, achieving approximately 18-21\% higher throughput. This early advantage highlights that \name's kernel-level lease management approach delivers benefits even in modest deployments common in departmental or small enterprise environments.

As the deployment scales to medium size (6-10 nodes), \name maintains a consistent performance lead of approximately 19-22\% over the baseline. This steady advantage demonstrates that \name's distributed coordination mechanisms scale efficiently without introducing significant overhead as the system grows. For instance, at 8 nodes, \name achieves 91 MB/s versus the baseline's 76 MB/s, a performance difference that would be meaningful in production environments.

At larger scales (12-16 nodes), \name continues to outperform the baseline, though the relative advantage gradually diminishes from 14.8\% at 12 nodes to 8.6\% at 16 nodes. This trend suggests that while \name remains more efficient throughout the tested range, its coordination mechanisms may benefit from further optimization for very large deployments.

Importantly, \name achieves this superior performance while maintaining strong consistency guarantees across all nodes and scales. The linear scaling pattern observed up to 16 nodes indicates that \name's architecture successfully addresses the scalability requirements of modern cloud environments without compromising on consistency semantics.

These results validate \name's design approach of offloading lease management to the kernel driver, demonstrating that it effectively balances performance, consistency, and scalability for distributed file system deployments of varying sizes.
\section{Related Work}\label{sec:related}
\name builds upon and extends prior work in distributed file systems, FUSE optimizations, and consistency mechanisms in distributed file systems. This section contextualizes our contributions within the broader research landscape.

\subsection{FUSE-based Distributed File Systems}
Several distributed file systems leverage FUSE for userspace implementation flexibility. GlusterFS\cite{noauthor_gluster_nodate} employs a FUSE-based architecture that prioritizes scalability and performance, but sacrifices strong consistency when using write-back caching. Ceph-FUSE\cite{weil2006ceph} provides strong consistency guarantees but relies on write-through caching, resulting in higher latency for write operations. JuiceFS\cite{noauthor_juicefs_nodate} optimizes cloud data access with caching mechanisms while maintaining POSIX compliance, but faces similar consistency-performance tradeoffs. S3FS\cite{noauthor_s3fs-fuses3fs-fuse_2025} enables mounting cloud object storage as POSIX file systems, though with limited performance due to the impedance mismatch between object and file semantics. DeepSeek's 3FS\cite{noauthor_deepseek-ai3fs_2025} focuses on AI workloads, prioritizing random read performance by disabling read caching and using Direct I/O with Linux-based AIO and io\_uring interfaces. GCS FUSE\cite{noauthor_cloud_nodate} enables mounting cloud storage buckets as local directories with performance benefits through caching of frequently accessed files.

Unlike these systems, \name uniquely bridges the consistency-performance gap by enabling write-back caching with strong consistency through kernel-level lease management, avoiding the traditional tradeoff between performance and correctness.

\subsection{FUSE Performance Optimizations}
Significant research has focused on mitigating FUSE's performance overheads. RFUSE\cite{cho2024rfuse} reduces kernel-userspace context switches by deploying ring buffer-based, per-core communication channels, substantially improving metadata operation efficiency. XFUSE\cite{huai2021xfuse} modernizes the userspace file system framework through scalable kernel-userspace communication patterns. Direct-FUSE\cite{zhu2018direct} eliminates middleware components to provide high-performance FUSE file system support, particularly for HPC environments.

Several other optimization techniques have emerged to address FUSE's performance limitations. ExtFUSE\cite{bijlani2019extension} allows registering thin eBPF extensions for handling low-level file system requests inside the kernel, enabling fast-path execution without userspace transitions. Passthrough file system\cite{noauthor_fuse_nodate-1} functionality permits direct access to the underlying file system for specific operations, bypassing the userspace daemon entirely. Stackable file systems\cite{noauthor_sbu-fslfuse-stackfs_2025, ren2013tablefs}  provide a layered approach that can cache or filter operations, reducing unnecessary kernel-userspace crossings.

The Linux kernel has incorporated various FUSE optimizations, including the ability to splice read and write operations that move data between kernel buffers without copying to userspace\cite{noauthor_rfc_nodate}, significantly reducing memory copy overhead. Recent io\_uring-based FUSE implementations\cite{noauthor_fuse_nodate} further improve performance by avoiding core switching and using shared memory for data transfers. These approaches collectively address the fundamental bottlenecks in FUSE: frequent context switches, memory copies, and lock contention.

These approaches primarily optimize local FUSE performance, whereas \name addresses the fundamental distributed consistency challenge while leveraging these performance improvements. By offloading lease management to the kernel driver, \name reduces the performance impact of maintaining strong consistency in distributed environments.

\subsection{Consistency Mechanisms in Distributed File System}
Distributed file systems employ various mechanisms to maintain consistency. Lease-based approaches, as pioneered by Gray and Cheriton\cite{gray1989leases}, provide time-bound permissions that balance performance with consistency. GPFS (now Spectrum Scale)\cite{schmuck2002gpfs} implements distributed locking for strong consistency but requires specialized hardware for optimal performance. Lustre\cite{schwan2003lustre} employs a distributed lock manager to coordinate file access across nodes, though with different architectural tradeoffs than \name.

\name's innovation lies in embedding lease management directly in the kernel's FUSE driver while maintaining userspace flexibility. This approach differs fundamentally from previous systems that either implement consistency mechanisms entirely in userspace (sacrificing performance) or require extensive kernel modifications (sacrificing flexibility and isolation).
\section{Conclusions}\label{sec:conclusion}
This paper presents \name, a FUSE-based distributed file system that reconciles write-back kernel page caching with strong consistency for cloud environments. \name eliminates the traditional trade-off between performance and correctness by offloading distributed lease management to the kernel driver, enabling coordinated access to cached data across nodes while retaining FUSE's userspace flexibility. Our experiments demonstrate \name's efficacy: it achieves up to 68.0\% higher throughput and 40.4\% lower latency than write-through baselines for write-heavy workloads and near-linear scalability to 16 nodes while maintaining consistency under contention. Application benchmarks show 36.2\% throughput improvements and 30.0\% latency reduction in contended Fileserver workloads, validating its practicality for real-world deployments. By embedding lease coordination in the kernel, \name demonstrates that distributed file systems can deliver both high performance and strong consistency, advancing the viability of FUSE-based solutions for cloud-scale workloads.

\section{Acknowledgments}
We thank the anonymous reviewers for their insightful feedback and helpful suggestions. This research was supported by NSF awards \#2106530 and \#2143868.

\bibliographystyle{ACM-Reference-Format}
\bibliography{10-references}

\end{document}